\documentclass[3p,twocolumn,a4paper,times,12pt]{elsarticle}

\usepackage{graphicx}
\usepackage{amssymb}
\usepackage{wasysym}   % diameter&arrow symbols.
\usepackage{makecell}
\usepackage{hyperref}

\newcommand{\ie}{\textit{i.\,e. }}
\newcommand{\eg}{\textit{e.\,g. }}
\newcommand{\etc}{\textit{etc }}

\journal{Atmosphere}

\begin{document}

\begin{frontmatter}
\title{Monitoring of Gamma Radiation Prior to Earthquakes at~a~Study of~Lithosphere-Atmosphere-Ionosphere Coupling in~Northern~Tien~Shan}

\author[iono]{N.\,Salikhov}
\author[lpi,iono]{A.\,Shepetov}\ead{ashep@www.tien-shan.org}
\author[iono]{G.\,Pak}
\author[iono]{S.\,Nurakynov}
\author[lpi]{V.\,Ryabov}
\author[kazgu,lpi]{N.\,Saduyev}
\author[itep,lpi]{T.\,Sadykov}
\author[iono]{Zh.\,Zhantayev}
\author[lpi]{V.\,Zhukov}

\address[iono]{Institute of Ionosphere, Almaty 050020, Kazakhstan}

\address[lpi]{P.\,N.\,Lebedev Physical Institute of the Russian Academy of Sciences (LPI), Leninsky pr., 53, Moscow, Russia, 119991}

\address[kazgu]{Al-Farabi Kazakh National University, Institute of Experimental and Theoretical Physics, Almaty 050040, Kazakhstan}

\address[itep]{Satbayev University, Institute of Physics and Technology, Almaty 050032, Kazakhstan}

\date{}

\begin{abstract}
Monitoring of radiation background in the near-surface atmosphere and of gamma rays, geoacoustic emission, and temperature in a borehole at 40\,m depth, as well as Doppler sounding on a low-inclined radio pass proceed at the Tien Shan mountain station (3340\,m a.s.l.) in Northern Tien Shan with common goal to search for seismogenic effects preceding earthquake. The flux of gamma rays in the borehole varies negligibly between the days, and it is not influenced by precipitations. Characteristic bay-like drops of the gamma ray flux were found (2--8)\,days before the M5.0--M6.2 earthquakes. In a M4.2 earthquake event with the 5.3\,km epicenter distance anomalies were detected (7--10)\,days before the earthquake in variation of the gamma ray flux, geoacoustic emission, and temperature. Simultaneously with gamma rays, a disturbance was detected in the Doppler shift of the ionospheric signal. Similarly, ionosphere disturbances triggered by the growth of radioactivity in the near-surface atmosphere were found at retrospective analysis of the Doppler shift data acquired after underground nuclear explosions at the Semipalatinsk testing site. This effect is considered from the viewpoint of the lithosphere-atmosphere-ionosphere coupling concept.
\end{abstract}

\begin{keyword}
earthquake\sep gamma radiation\sep geoacoustic\sep lithosphere-atmosphere-ionosphere coupling\sep Doppler sounding\sep nuclear explosion
\end{keyword}

\end{frontmatter}

\section{Introduction}
In spite of many years of active study of the anomalous phenomena observed in geophysical fields before earthquake, the forecast problem  of powerful earthquakes still remains rather urgent and demands special attention. One of the most significant markers of approaching earthquake is the intensification of the release from rock of the radioactive gases $^{220}$Rn, $^{222}$Rn and their short-lived daughter products under stimulating influence of the change of strained lithosphere medium during the period of earthquake preparation. This phenomenon was reported of in many investigation works which were accomplished in a variety of seismically active regions \cite{introref1,introref2,introref3,introref7, introref4,introref5}. According to \cite{introref6}, in the seismic region of Northern Tien Shan usually an earthquake preparation process starts to reveal itself among the radon exhalation data 3--4 months afore the seismic event, and is especially clear manifested 1--2 weeks before the earthquake.

For detection of the anomalies of radon emanation of seismic origin it is convenient to use the registration of the intensity of gamma radiation emitted by the radioactive nuclei-products of the $^{222}$Rn decay \cite{introguili}. The method of gamma radiation monitoring in the energy range of tens of keV---a few MeV has shown itself as an effective indirect tool for monitoring of the current radon content and timely detection of the indications of possible earthquake. For example, a significant increase of the level of radiation background was observed few days before the M5.2 and M6.7 earthquakes at Northern Taiwan \cite{introref13}. Anomalous rise of gamma ray intensity was detected also some days before the local earthquake swarms at the East of Taiwan in March and May 2014, and during the two weeks preceding the M5.9 Fangly earthquake. After exclusion of possible temperature effect the authors  \cite{introref10} suppose the prognostic significance of these radiation increases. In  \cite{introref14} it is shown that a single strong earthquake or a swarm of close local earthquakes can influence the gamma ray background even outside of seismically active region.

Nevertheless, in a number of studies, such as \cite{introref7,introref8,introref9}, it was stressed the selection difficulty of the impact of the processes of seismological origin on the radon concentration among the multitude of different effects of non-tectonic nature, such as soil humidity, precipitation, \etc. The interfering effects can cause fluctuations of radon content in the near-surface atmosphere similar to the earthquake precursors, or mask the real release of radon from the depth of lithosphere which can be restored only through a special mathematic procedure \cite{introref11}. As illustration, in \cite{introref12} the dependence of radon emanation on the weather was found even at the level $\sim$10\,m deep under the surface of the ground. According to \cite{introref14a}, the majority of anomalies observed during pre-earthquake period in the records of gamma ray intensity might be connected with meteorological, hydrological, and other environmental phenomena instead of seismic processes, since the radiation time series reflect a wide spectrum of various physical effects. Direct influence of the weather and other local factors on the intensity of radiation background was demonstrated in many publications, \eg \cite{introref12,introref14,rainsour2009,introref16,introref17}. Consequently, processing of the prolonged time series of radon concentration measurements demands application of specific mathematical methods \cite{introref10,introref11,introref16}.

% doppler
The literature data of last decades convince in the necessity of comprehensive monitoring of the parameters of geophysical environment in order to prove or disprove the involvement of identified anomalies in the earthquake preparation process. An increased attention is paid to the study of the relationship between the dynamics of lithospheric processes and the appearance of disturbances in the ionosphere which precede large earthquakes \cite{introref19,introref21,introref202,introref201}. Comprehensive monitoring makes it possible to trace the relationship between the deformation of the earth's crust, radon exhalation, and gamma radiation. The rise of radiation background in the near-surface atmosphere leads to ionization of the atmospheric boundary layer, which, according to the authors of the li\-tho\-sphe\-re-at\-mo\-sphe\-re-iono\-sphe\-re coupling concept \cite{introref18,introref22,pulinets2015} is one of initial links in the chain of disturbance propagation from the lithosphere up to the ionosphere heights.

%Thus, in the period of earthquake preparation it can be existing united process of a consecutive transfer of various disturbances through the different geophysical media: the lithosphere, atmosphere, and ionosphere, which seems rather promising for a more unambiguous selection of the indicators of increasing seismic activity.
% 
% re-placed from Conclusion:
%Doubtless, simultaneous detection of the anomalies at once in a number of different geophysical media permits to make more reliable conclusion on their seismological nature than sporadic observation of some peculiarities in behaviour of separate environmental parameters, each of which undergoes the influence of a variety of many side effects.
%
%As well, from an experimentalist's point of view the detection of ionosphere anomalies may occur more advantageous then the direct monitoring of the level of radioactivity at the earth's surface, since it permits to avoid essentially the interfering influence of the local weather, humidity, \etc on the signal of interest.

There exist a lot of appropriate methods for investigation of seismo-ionospheric disturbances among which the radiophysics method of Doppler ionosphere sounding is one of most effective \cite{introref18,introref22,pulinets2015,introref29,introref27,introrefhayakowa}. The method of Doppler sounding at a low-inclined radio pass is used also in the present work \cite{dopplernazyf1}.

%One of the most effective techniques to study the ionosphere disturbances is the method of Doppler shift measurement by radio-sounding of the ionosphere layers \cite{introref27,introref29}. %introref28,

An important condition for research of seismogenic anomalies in geophysical fields is localization of the measuring equipment near the sources of strong earthquakes. Such is the geographical region of the mountain ridges Zailiysky Alatau and Kungey Alatau in Northern Tien Shan which encompasses the sources of several disastrous earthquakes: the Vernen (M7.3, 1887), Chilick (M8.3, 1889), and Kemin (M8.2, 1911) ones. Here reside an experimental base of the Institute of Ionosphere, the ``Orbita'' radio-polygon (N43.05831,\,E76.97361; 2750\,m above the sea level), and the Tien Shan mountain scientific station of the Lebedev Physical Institute (N43.03519,\,E76.94139; 3340\,m a.s.l.).
%With account to the fundamental seismological law of earthquake reiteration in a seismically active area, such disposition was a premise for development at the both scientific centers of an investigation direction aimed to the search for possible indicators of approaching earthquakes.

%Of no small importance is also the closeness of the large inhabited localities to the mentioned earthquake sources and the currently observed intensification of seismic processes in this region, which determined the necessity of complex study of the parameters of geophysical fields at the final stage of earthquake preparation with a general purpose to trace the detected anomalous effects from their generation in the seismically active lithosphere areas up to the ionosphere.

Another prerequisite for the said research activity was the existence at the Tien Shan mountain station of a multipurpose experimental complex which includes the detectors of high-energy cosmic rays, the gamma- and neutron detectors, a weather station, % the sensors of electromagnetic field,
acoustic receivers, \etc, well suitable for prolonged multichannel monitoring of the surrounding environmental background \cite{thunderour2016,thunderour2021iskra,undgacou2019acts}. Also, among other positive factors is the unique location of the Doppler ionosonde at the ``Orbita'' radio-polygon which makes  possible the ionosphere study within the distance of up to (200--250)\,km from the radio-wave projection point, including the sources of the Vernen and Kemin earthquakes.
%
%Besides, a special Doppler sounding equipment situated at the ``Orbita'' polygon can be used for continuous check of the current ionosphere state over an extended seismically active mountain region of Tien Shan, Pamir and Himalayas \cite{dopplernazyf1,dopplernazyf2}.
%
%The main goal of using such heterogeneous combination of various detection methods is to trace, as detailed as possible, the propagation of anomalous disturbances of seismic origin through the different geophysical media, and, by comparison of these anomalies, to select possibly unambiguously the imprint of a preparing earthquake in the time series of the registered physical values against the bulk of interfering side effects. Further on, such observations may be applied for elaboration of the practically useful earthquake forecast methods.
%
The main goal of using such complex combination of investigation methods is to track consistently the propagation in geophysical media of anomalous disturbances, of a presumably seismic origin, from the lithosphere and up to the lithosphere heights. 

Taking into account rare recurrence of powerful earthquakes, a rather prolongate observation period is needed to confidently verify the connection of any detected anomaly with the preparation process of a earthquake. 
%There exists also a supposition that some anomalies can reveal themselves only if the magnitude of earthquake exceeds some threshold.
This circumstance stipulates the necessity of continuous monitoring of various physical parameters in the surrounding environment, which at the Tien Shan station has been nearly uninterruptedly proceeding during the last decade.

The subject of current publication is to present the results of a complex monitoring of gamma radiation intensity, both in the near-surface atmosphere and at the depth of 40\,m in a deep borehole, as well as the Doppler shift measurements of radio wave reflected from the ionosphere. The data presented here were acquired on the eve of several earthquakes. Also, the possibility is discussed of the propagation of the revealed seismogenic disturbances over the chain of li\-tho\-sphe\-re-at\-mo\-sphe\-re-iono\-sphe\-re coupling.

\section{Experimental technique}
\label{sectitechni}

\paragraph{Detectors of the gamma radiation background}
For monitoring of the intensity of gamma radiation in the surrounding environment of the Tien Shan mountain station the scintillation detectors are used, the sensitive part of which consists of an inorganic NaI(Tl) crystal optically coupled with a photomultiplier tube. Such scintillators have a rather high efficiency, of the order of (20--80)\%, of gamma ray detection in the energy range of tens of keV---a few MeV. 

By registration of radiation intensity the gamma detectors operate in the pulse counting mode. From the anode output of the photomultiplier tube the electric pulses with the amplitudes proportional to the energy of gamma ray quanta absorbed in the scintillator come to a set of amplitude discriminators with monotonously increasing operation thresholds, which ensure separate registration of gamma ray signals at once in a number of energy diapasons, starting from 30\,keV and up to (1.5--2)\,MeV. For the each energy range the procedure of intensity measurement consists of counting, by the means of a digital scaler scheme, of the shaped pulses which have come from the output of corresponding discriminator during a fixed exposition time. Typically, the length of exposition in  monitoring experiments at the Tien Shan station equals to 10\,s. As it has revealed in the course of the considered study, such a rather high temporal resolution of the intensity measurement was crucial for disclosure of the weak and short time variations in radiation background.

An absolute energy calibration of discriminator thresholds was done using the etalon sources of gamma radiation made on the basis of $^{241}$Am, $^{22}$Na, $^{60}$Co, and $^{137}$Cs radioactive isotopes.

Of the two gamma radiation detectors used in the considered experiment the one having the sizes of scintillator crystal of ($\diameter$150$\times$110)\,mm$^2$ was installed inside a wooden building under a light 1\,mm thick iron roof, where the temperature varied between the limits of (10--14)$^\circ$C in dependence of the year season. Another one, with the sizes ($\diameter$40$\times$40)\,mm$^2$, operated under the surface of the ground at a depth of $\sim$40\,m in a dry borehole (see below). % with excessively stable temperature conditions.

%The background flux of thermal neutrons at the Tien Shan station is measured with the use of neutron detectors built on the basis of cylindrical ($\diameter$30$\times$1000)\,mm$^2$ gas discharge ionization counters with a special gas filling which includes the $^3$He isotope. When a low-energy neutron comes inside such counter the $n$($^3$He,$^3$H)$p$ nuclear reaction takes place, and the charged products of the latter are detected by the counter as an electrical pulse at its anode. Determination of the integral intensity of neutron flux is made, again, by counting these pulses during a fixed exposition time.

%Two detectors of the thermal neutron flux the data of which will be presented below consisted of aluminum boxes with six neutron counters placed in the each. One of them (the detector \#1) was placed close to the surface of the ground inside a light wooden shelter, another one (detector \#2)---on the second floor of a capital building with concrete floors and iron roof.

More detailed description of the detectors applied at the Tien Shan mountain station for monitoring of the radiation background of various kinds in the surrounding environment can be found in \cite{thunderour2016,thunderour2020touritge}.

\paragraph{The underground detectors in a deep borehole}
At the territory of the Tien Shan mountain station, at an altitude of 3340\,m a.s.l., there exists a borehole which it is convenient to use for investigation of seismogenic effects. The total depth of the borehole is of about 300\,m, its bottom part is filled with water up to level of $\sim$150\,m.
There are four digital thermometers with automatic temperature recording installed both at the surface of the ground close to the borehole and inside the latter, at the levels of 1\,m, 25\,m, and 40\,m. %As it was found, at the depth of 25\,m and below the temperature remains nearly constant during the whole year and oscillates only within the tight limits of $\sim$0.01$^\circ$C.

As well, a gamma detector with a small-sized, ($\diameter$40$\times$40)\,mm$^2$, scintillator was installed in the borehole at the depth of 40\,m with the purpose to monitor the intensity of gamma radiation in the subsoil layers of rock. It was found that  at this depth the temperature remains nearly constant during the whole year and equals to (2.500$\pm$0.001)\,$^\circ$C. Thus, the temperature conditions inside the borehole ensure stable operation of both  the scintillation detector and its electronic equipment. Besides temperature stability, such disposition of the detector deep underground permits to avoid the influence of local weather on the measured radiation background.

The counting rate registration of the signals from the gamma detector placed in the borehole is made quite in the same way as for the detectors operating at the surface of the ground. The analog discriminator part of the electronic equipment is hosted in a light cabin near the upper end of the borehole and is kept there under stabilized temperature conditions of (20$\pm$0.25)\,$^{\circ}$, independently on the season of year. Transmission of the analog pulse signals from the depth of the detector location up to  the discriminators board is made over a pair of twisted wires connected to an operational amplifier with differential inputs.

Another experimental equipment installed in the borehole is a microphone for continuous monitoring of the level of acoustic noises of presumably seismic origin. The microphone, with a sensitivity of 25\,mV/Pa in the frequency range of (0.5--10)\,kHz, was placed at the depth of 50 m. Its analogue output signal is transmitted to the surface via a symmetric connection line made of a pair of twisted wires. Operation of the signal is made by an integrated circuitry of 12-bit analog-to-digital conversion (ADC) placed in a cabin which is installed near the upper opening of the borehole. With account of the microphone sensitivity parameter of 25\,mV/Pa and the amplification coefficient $K\approx 100$ of the analogue electronic tract which stands before the ADC input, the ADC code $A\approx 100$ corresponds to the absolute value of acoustic pressure $p$ of about 0.05\,Pa (3\,dB) in the point of the microphone disposition, while $A\approx 1000$ corresponds to $p\approx 0.5$\,Pa (4\,dB). The highest possible code for 12-bit conversion, $A=4095$, corresponds to the pressure of $p\approx 2$\,Pa (5\,dB).

All the measurement time the digitization of the microphone signal was uninterruptedly going on with a moderate periodicity of 1\,ms (1\,kHz). To agree more adequately the digitization period with the range of the microphone sensitivity, an integral envelope of the signal was detected immediately during the measurements and recorded along the original signal. This detection was made by an active low-pass analogue filter which was included into the electronic tract of signal operation. The integration time constant of the filter circuit equals to 7\,ms, \ie to few periods of the original signal with its kHz order frequency.

More detailed description of the acoustic detector is given in \cite{undgacou2019acts,undgacou2018}.

\begin{figure}
{\centering
\includegraphics[width=0.5\textwidth, trim=0mm 0mm 0mm 0mm]{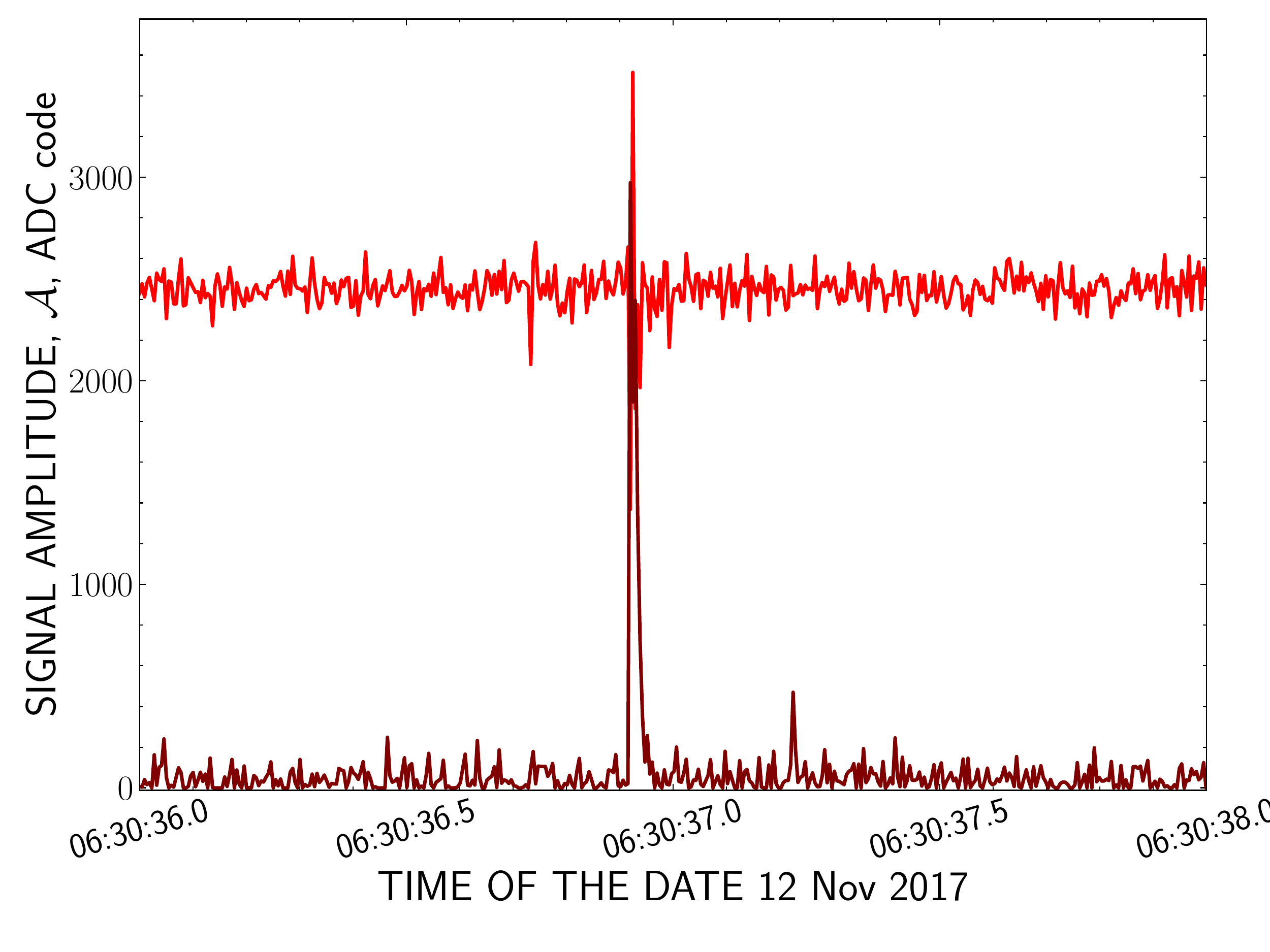}
\caption{Sample record of the acoustic detector data: the upper distribution is the original microphone signal, the lower one---its smoothed envelope detected by the low-pass filter. The short-time peaks are seen in the envelope distribution.}
\label{figiacoupeak}}
\end{figure}

Practically, it was convenient to use as characteristic measure of the current level of acoustic noises the amount of the short-time, $\sim$(10--15)\,ms, outburst peaks with an amplitude above some pre-defined threshold, $A\geqslant A^\star$, which have been detected in the record of the envelope signal during a fixed period of time (typically, one minute). A sample of such record of the acoustic signal is shown in Figure\,\ref{figiacoupeak}; the threshold of ADC code $A^\star$ for the peak counting procedure in the considered experiment was accepted to be equal to 300.
% sample illustration: 
% ~/E/k09007 -b'acou' -t'nev' -n'1:2500|2:0' -y'0|4000' -d'28.11.2021'
% 28.11.21 00:16:14 ... 28.11.21 00:20:24

\paragraph{The Doppler sounding equipment}
The search for an ionospheric response to activation of seismic processes in the considered experiment is made with the use of a hard- and software complex for Doppler measurements of the ionosphere signal on a low-inclined radio-pass \cite{dopplernazyf1}. The radio-receiving part of the complex is placed at the radio-polygon ``Orbita'' of the Institute of Ionosphere, 2.6\,km apart from the detector system of the Tien Shan mountain station. The transmitter is hosted at a distance of 16\,km from the receiver and is separated from the latter by a mountain ridge. As well, the signals of some far broadcasting radio-stations located at the distances of up to 4000\,km were used for the purpose. The frequency range of the radio-complex operation is (3--10)\,MHz; switching of the working frequency is made under the control of an automatic program in dependency on the day time and year season.

The measurement of the Doppler shift frequency of radio wave reflected from the ionosphere is based on the phase-locked loop (PLL) operation principle which permits to convert the Doppler frequency into proportional voltage at the output of phase detector. The width of the holdoff band of the PLL loop applied in the current experiment is 15\,Hz, and the non-linearity of its frequency conversion characteristic does not exceed 0.46\%, which is quite sufficient for the high-quality measurements of the ionospheric Doppler shift.

%Доплеровский ионозонд может работать в диапазоне частот от %1000 кГц до 30000 кГц (1 МГц - 30 МГц). Шаг перестройки частоты  радиоприемного устройства 1 Гц.
%Радиочастоты своего передатчика: 
%5121 кГц  — дневная частота
%2963 кГц  — ночная 
%Используем также частоты радиовещательных станций:
%4010 кГц Красная речка Кыргызская республика 
%4850 кГц КНР Урумчи
%5860 кГц Кувейт
%7205 кГц КНР Kashi-Saibagh
%7245 кГц Таджикистан
%7275 кГц КНР Урумчи
%7570 Северная Корея
%9560 кГц  КНР Урумчи

\section{Experimental data}

\subsection{The variation of the background flux of gamma rays in the subsoil rock layers and near-surface atmosphere}

\begin{figure}
{\centering
\includegraphics[width=0.5\textwidth, trim=0mm 0mm 0mm 0mm]{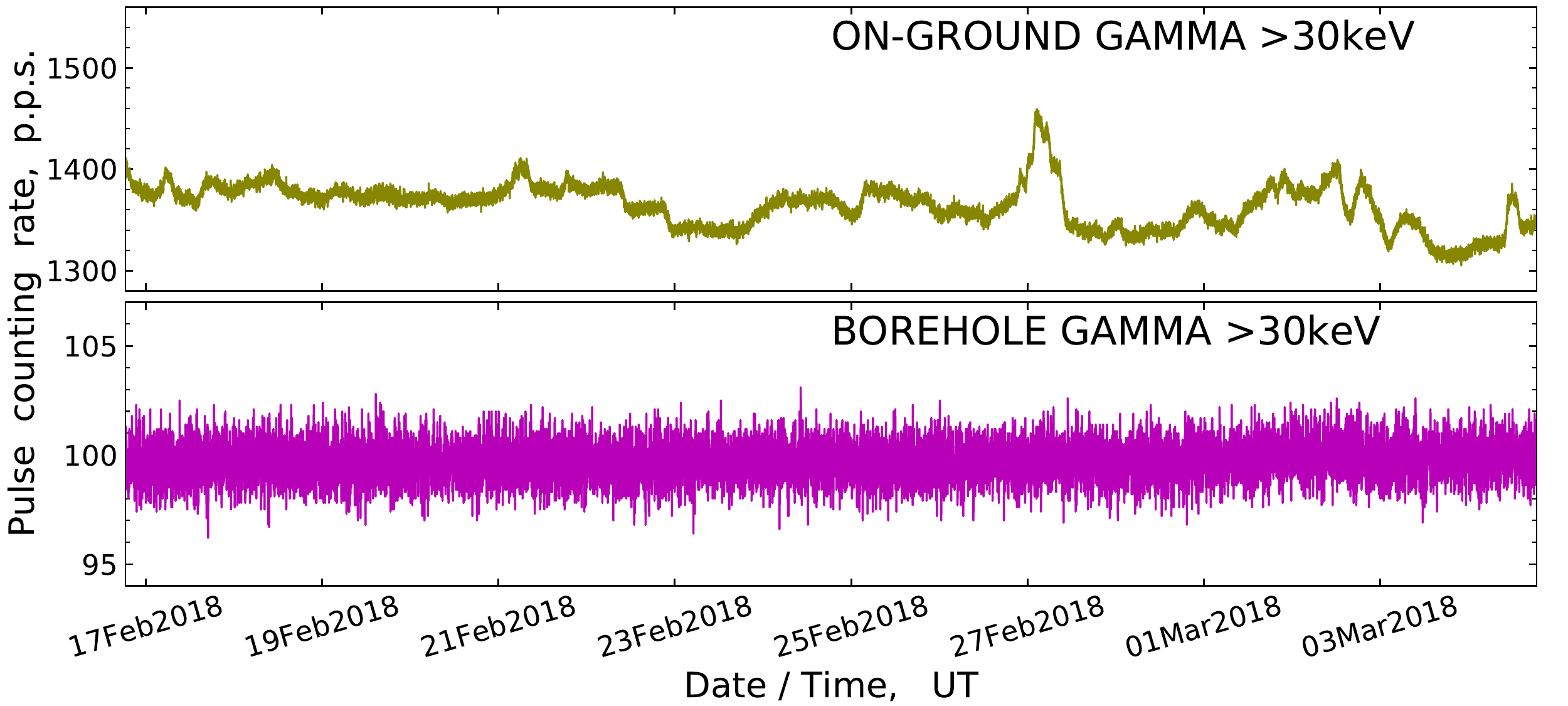}
\includegraphics[width=0.5\textwidth, trim=0mm 0mm 0mm 0mm]{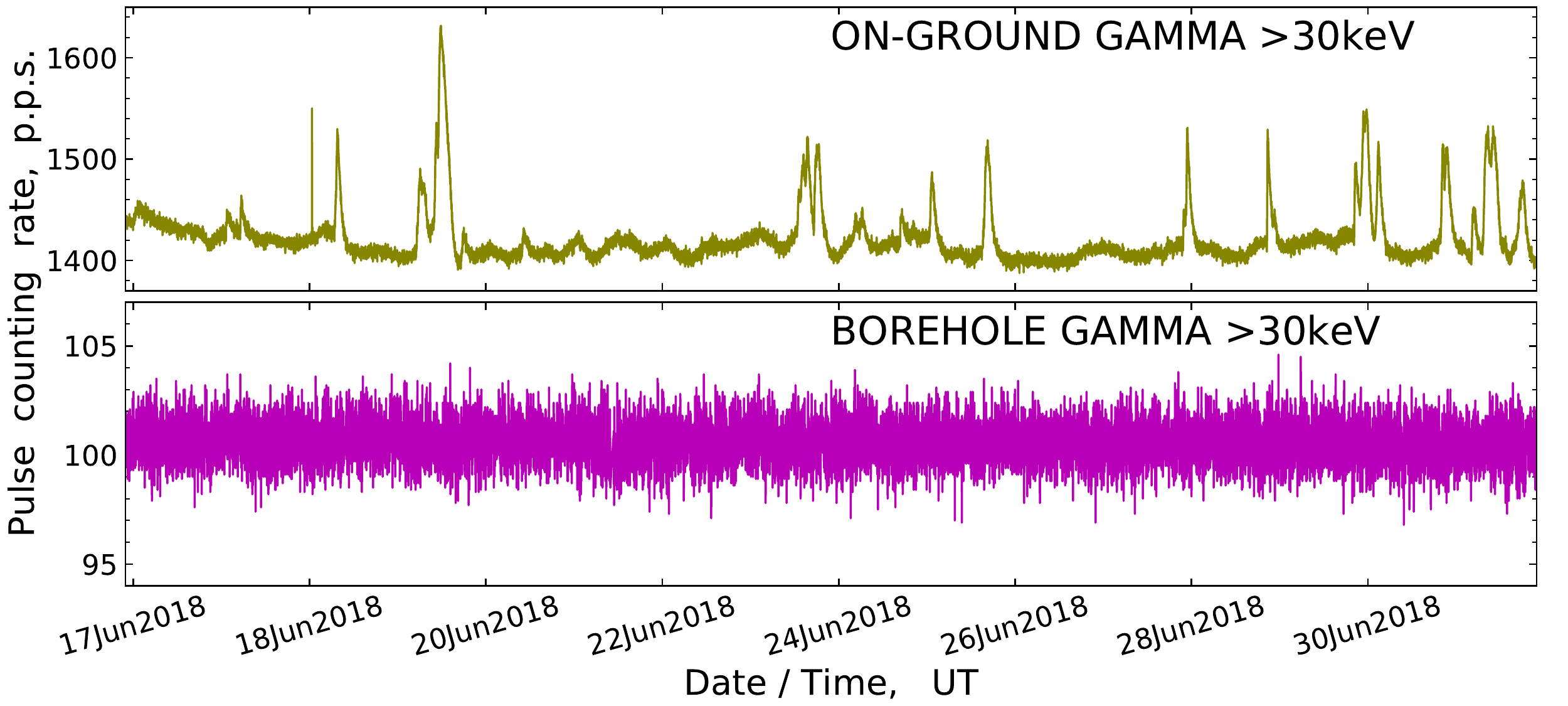}
\caption{Comparison of the radiation intensity measurements which were simultaneously done by the gamma detectors installed above the surface of the ground and in the borehole. Upper frame: a 17\,days long period randomly taken in a winter season. Lower frame: similar comparison for a summer season. All intensity values are expressed in the amount of pulse signals gained per one second (p.p.s.) in corresponding energy range.}
\label{figigammomoc}}
\end{figure}

Figure\,\ref{figigammomoc} demonstrates a sample of typical monitoring data of the intensity of gamma radiation as they were measured at the Tien Shan station by the detectors installed at the surface of the ground and at the depth of 40\,m in the borehole. As it is explained in Section\,\ref{sectitechni}, the monitoring of the intensity of gamma ray signals in both detectors was done with 10\,s periodicity, and for a number of energy thresholds of registered gamma ray quanta. The data plotted in Figure\,\ref{figigammomoc} relate to the range with the minimum energy threshold of registered gamma rays, $E_\gamma \geqslant$30\,keV, and all intensity values there are expressed in the units of the counting rate, \ie they are normalized to the amount of the pulse signals which come per one second (p.p.s.) in a given energy range of the corresponding detector.

There is an essential difference in Figure\,\ref{figigammomoc}  between the time histories of gamma ray intensity recorded in the near-surface atmosphere by the surface detector, and the measurements of the subsoil radiation background in the borehole. Firstly, the intensity of gamma rays above the surface of the ground varies in noticeably wider limits than that in the borehole: if the relative variation of the counting rate values $(I_{max}-I_{min})/I_{mean}$ in the atmosphere equals to $\sim$11\% in the winter season and $\sim$17\% in summer, in the underground environment it remains about 8\% independently on the time of year. Besides random fluctuation, in the time series of the surface detector it can be seen also a quasi-sinusoidal oscillation with a period duration close to one day. 

Secondly, an appreciable  difference exists between the average levels of the radiation background measured in the atmosphere by the surface detector in winter, $I_{mean}=(1365\pm 23)$\,s$^{-1}$, and in summer, $I_{mean}=(1422\pm 28)$\,s$^{-1}$. In contrast, corresponding estimations for the counting rate data of the detector installed in the borehole are practically same, $(99.7\pm 0.9)$\,s$^{-1}$ and $(101\pm 1)$\,s$^{-1}$, \ie their difference remains in the limits of statistical fluctuation. 
% on-ground - winter: (1459-1305)/1364 = 11%
% on-ground - summer: (1631-1387)/1422 = 17%
% underground: (104.6-96.80)/100.59=8%
%

%\begin{figure*}
%{\centering
%\includegraphics[width=\textwidth, trim=0mm 0mm 0mm 0mm]{backgra.pdf}
%\caption{A sample of the data of radiation monitoring with different energy thresholds in the borehole.}
%\label{figigammomoa}}
%\end{figure*}

At last, multiple short-time peaks of the counting rate with  relative amplitude of up to a few tens of percents and typical duration of (1--2)\,h can be seen in the plots of the surface detector data, especially during the summer season. These are the imprints of precipitations fallout in the form of snow, rain, or hail, which caused temporary raise of the local radioactive background in the near-surface atmosphere \cite{rainsour2009}, but practically did not exert any influence on the intensity of the subsoil gamma rays measured by the borehole detector.

%Excellent stability of the radiation intensity under the soil surface reveals itself also in the different energy ranges. In Figure\,\ref{figigammomoa} the prolonged time series of the counting rate in the borehole detector are shown simultaneously for the four thresholds of registered gamma ray quanta, 30\,keV, 150\,keV, 250\,keV, and 500\,keV correspondingly. All these distributions similarly demonstrate an uniformly stable behaviour.

Thus, it can be stated that for the search of such feeble effect as the anomalies specific for the times of approaching earthquake a proper method may be the monitoring of the subsoil radiation background in the energy range of a few tens---hundreds of keV, at the depth of a few tens of meters under the surface of the ground. The data on radiation intensity thus obtained are free from the influence of local weather, as well as of diurnal and seasonal effects. A proper device for this purpose may be the underground gamma detector installed at the depth of 40\,m in the borehole of the Tien Shan mountain station.

\subsection{Peculiarities of radiation background in the times of earthquake preparation}
\label{sectisinu}

Up to the present time, the monitoring of geophysical parameters at the Tien Shan mountain station has been continuing about a decade. During this period there were observed 7~earthquake events with an epicenter in the limits of Dobrovolsky radius \cite{dobrovolsky} relative to position of the observation point, \ie the station together with its detectors has occurred within the zone of earth's crust deformation connected with the earthquake. All these events are listed in Table\,\ref{tabitabi}. Four leading columns of this table contain designations of the earthquake time and epicenter coordinates; the magnitude $M$ and the depth of the earthquake focus \cite{seismo_some}; an estimation of the Dobrovolsky radius calculated as $R_D=10^{0.43 M}$\,km; and the distance from the epicenter point to the borehole.

\begin{table*}
\setcellgapes{2pt} % both defined
\makegapedcells    % in 'makecell'
\renewcommand{\thempfootnote}{\arabic{mpfootnote}}
\begin{minipage}{\textwidth}
\begin{center}
\caption{The list of the earthquakes with anomalous effects detected during the period of their preparation in the time series of gamma ray intensity.}
\label{tabitabi}
\begin{tabular*}{\columnwidth}{@{\extracolsep{\fill}}c|c|c|c|c|c|c}
\hline
\hline
\makecell{time, UT;\\epicenter\\coordinates} &
\makecell{magnitude,\\$M$;\\depth,\\km} &
\makecell{Dobro-\\volsky\\radius,\\$R_D$,\\km} &
\makecell{distance\\to the\\borehole,\\km} &
\makecell{effect\\in the\\bore-\\hole} &
\makecell{effect\\in the\\atmo-\\sphere} &
\makecell{days\\between\\the effect\\and the\\earthquake}
\\
\hline
\hline

% 1
\makecell{1 May 2011\\02:31:27;\\N43.71 E77.66}&
\makecell{5.8;\\6}&
\makecell{312}& 
\makecell{92}&
\makecell{no data\footnote{The detectors in the borehole were installed only at the end of the year 2017.}}&
\makecell{\Large{+}}&
\makecell{2}
\\
\hline

% 2
\makecell{15 Mar 2015\\14:01:00;\\N42.95 E76.92}&
\makecell{5.2;\\20}&
\makecell{172}&
\makecell{15}&
\makecell{no data\footnotemark[1]}&
\makecell{\Large{+}}&
\makecell{6}
\\
\hline

% 3
\makecell{30 Dec 2017\\15:55:46;\\N43.10 E77.90}&
\makecell{4.2;\\10}&
\makecell{64}&
\makecell{5}&
\makecell{\Large{+}}&
\makecell{\Large{+}}&
\makecell{7}
\\
\hline

% 4
\makecell{31 Aug 2018\\21:21:27;\\N43.02 E77.48}&
\makecell{5.0;\\25}&
\makecell{141}&
\makecell{51}&
\makecell{\Large{+}}&
\makecell{\Large{+}}&
\makecell{2}
\\
\hline

% 5
\makecell{14 Sep 2018\\22:15:03;\\N41.09 E77.22}&
\makecell{5.3;\\5}&
\makecell{190}&
\makecell{127}&
\makecell{\Large{+}}&
\makecell{\Large{+}}&
\makecell{6}
\\
\hline

% 6
\makecell{19 Jan 2020\\13:27:57;\\N39.90 E77.18}&
\makecell{6.2;\\10}&
\makecell{463}&
\makecell{354}&
\makecell{\Large{+}}&
\makecell{\Large{--}}&
\makecell{8}
\\
\hline

% 7
\makecell{15 Feb 2020\\03:31:53;\\N41.88 E79.22}&
\makecell{5.6;\\5}&
\makecell{256}&
\makecell{229}&
\makecell{\Large{+}}&
\makecell{\Large{+}}&
\makecell{7}
\\
\hline

% 8
%\makecell{26 May 2022\\15:00:36;\\N42.87 E77.66}&
%\makecell{5.5;\\10}&
%\makecell{232}&
%\makecell{60}&
%\makecell{\Large{+}}&
%\makecell{\Large{--}}&
%\makecell{10}
%\\
%\hline

%\hline
\end{tabular*}
\end{center}
\end{minipage}
\end{table*}

At the time of the earthquakes mentioned in Table\,\ref{tabitabi} continuous monitoring of the ionizing radiation background at the Tien Shan station was going on, either with the use of the gamma ray detector installed above the surface of the soil, or with the one put into the borehole, or with both (see Section\,\ref{sectitechni}).

As it is seen in the plots of Figure\,\ref{figigammomoc},
% and \,\ref{figigammomoa},
generally the original results of gamma intensity monitoring constitute a time series type distribution in which it hardly can be found any perceivable deviation above the usual statistic fluctuations and imprints of various interfering effects. Thus, a need exists in a special analysis procedure for searching amongst these data for any unusual effects which could appear in the time periods preceding earthquake. In the course of operation of the monitoring data gained in the considered experiment such procedure has been elaborated indeed. Next on, this procedure is explained in detail on an example of the January~19, 2020 earthquake event.

\begin{figure}
{\centering
\includegraphics[width=0.5\textwidth, trim=0mm 30mm 0mm 30mm]{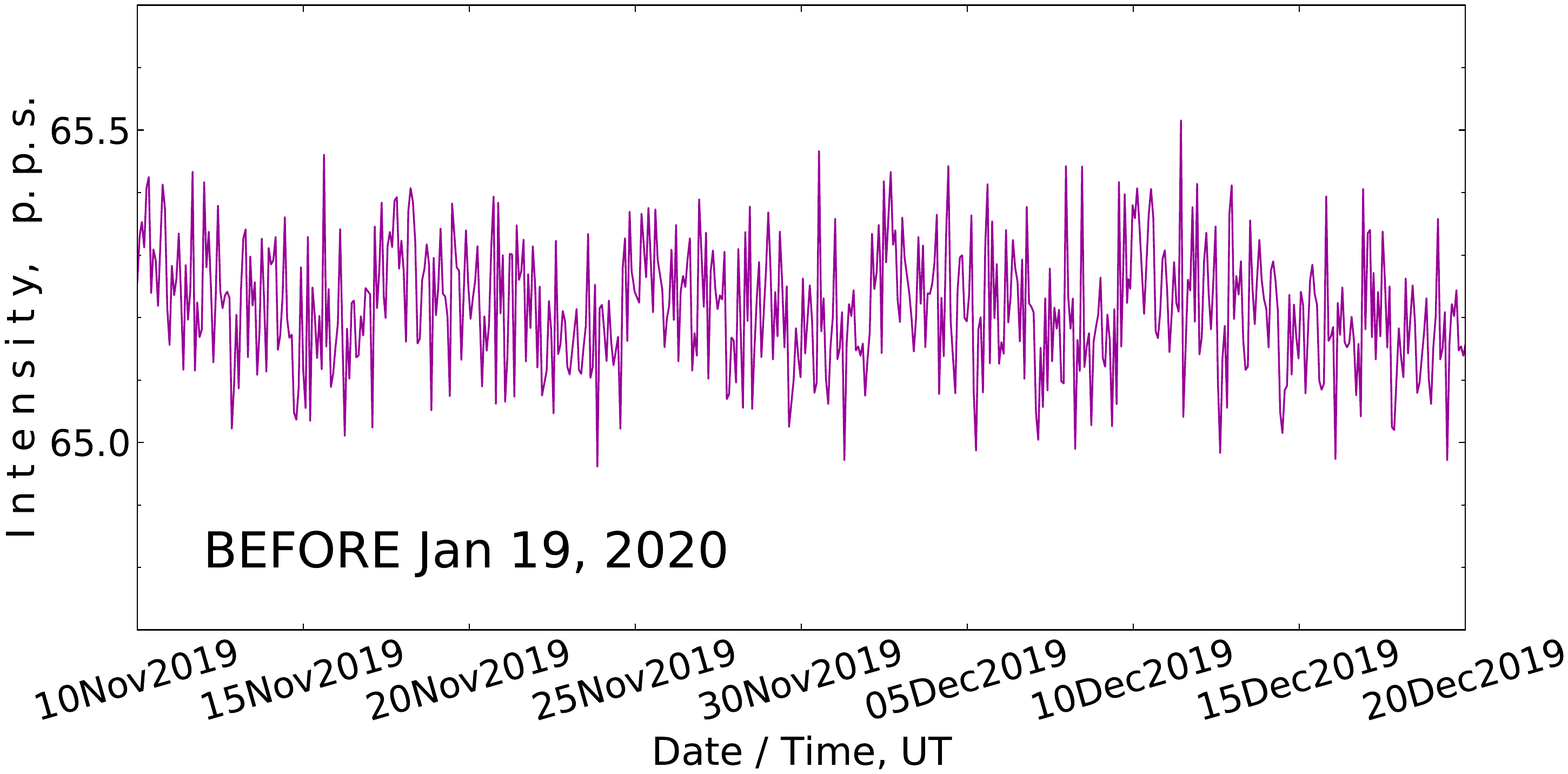}
\includegraphics[width=0.5\textwidth, trim=0mm 30mm 0mm 30mm]{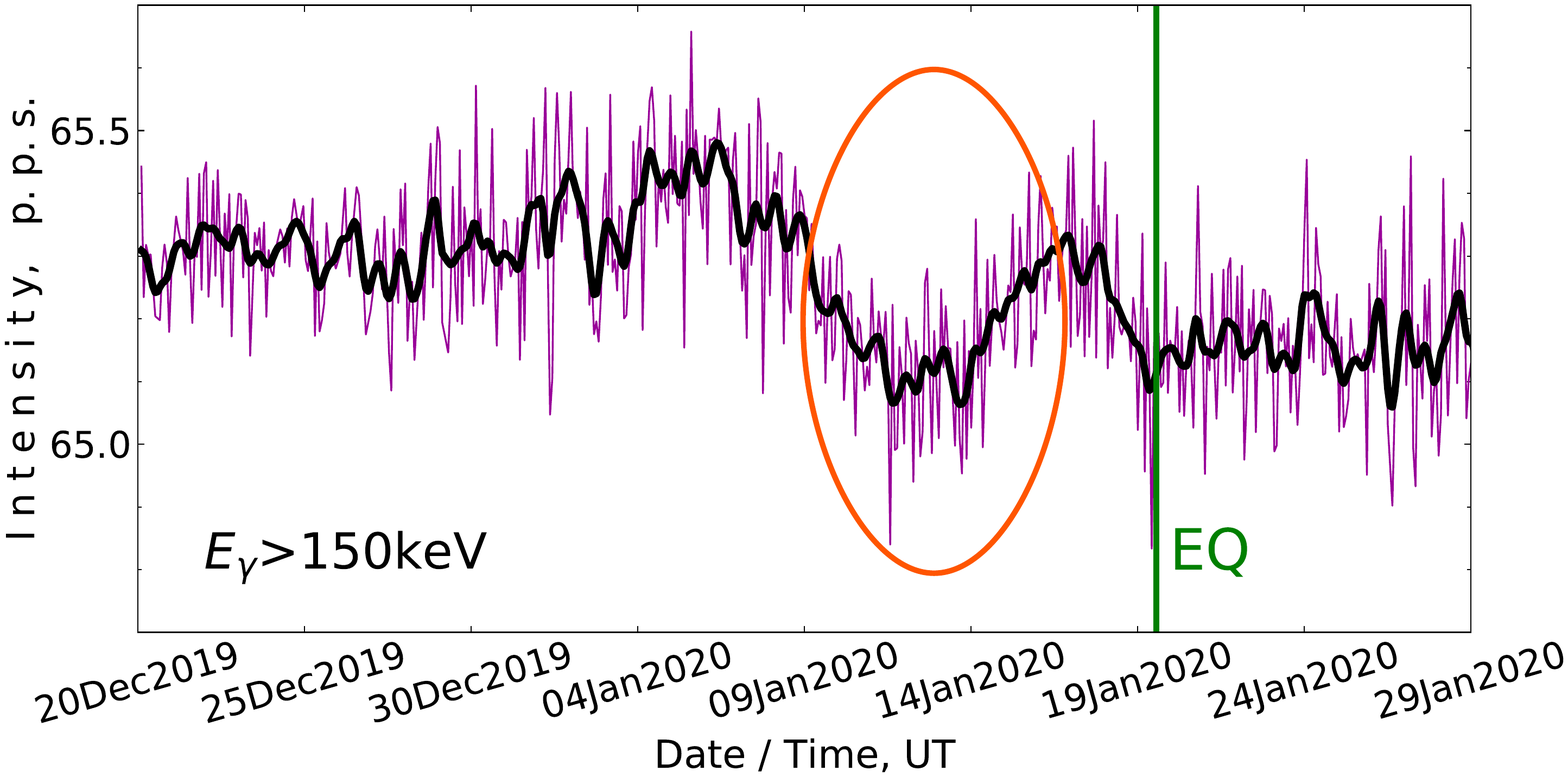}
\includegraphics[width=0.5\textwidth, trim=0mm 30mm 0mm 30mm]{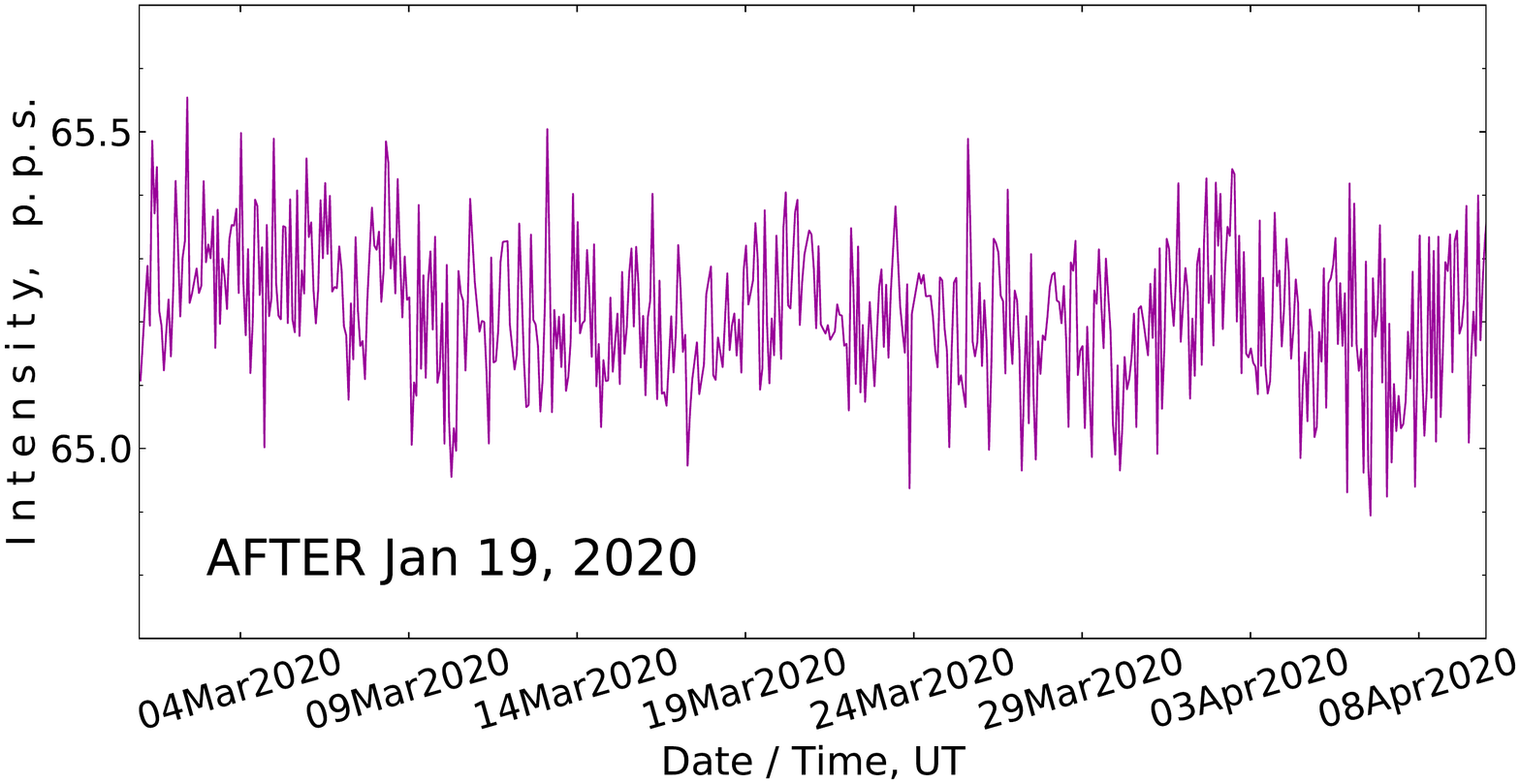}
\caption{
Time distribution of the intensity of gamma ray signals registered in the second channel of the borehole detector (that with the minimum gamma ray energy 150\,keV) after application of the averaging and filtering procedure (see text). Upper panel: the time period before the January~19, 2020 earthquake; middle panel: the period of the earthquake; bottom panel: the period after the earthquake. The moment of the earthquake (EQ) is marked by the vertical line, and the period of the preceding bay-like drop in the intensity curve by an ellipse.
}
\label{figigammomob}}
\end{figure}

\begin{figure}
{\centering
\includegraphics[width=0.5\textwidth, trim=0mm 0mm 0mm 0mm]{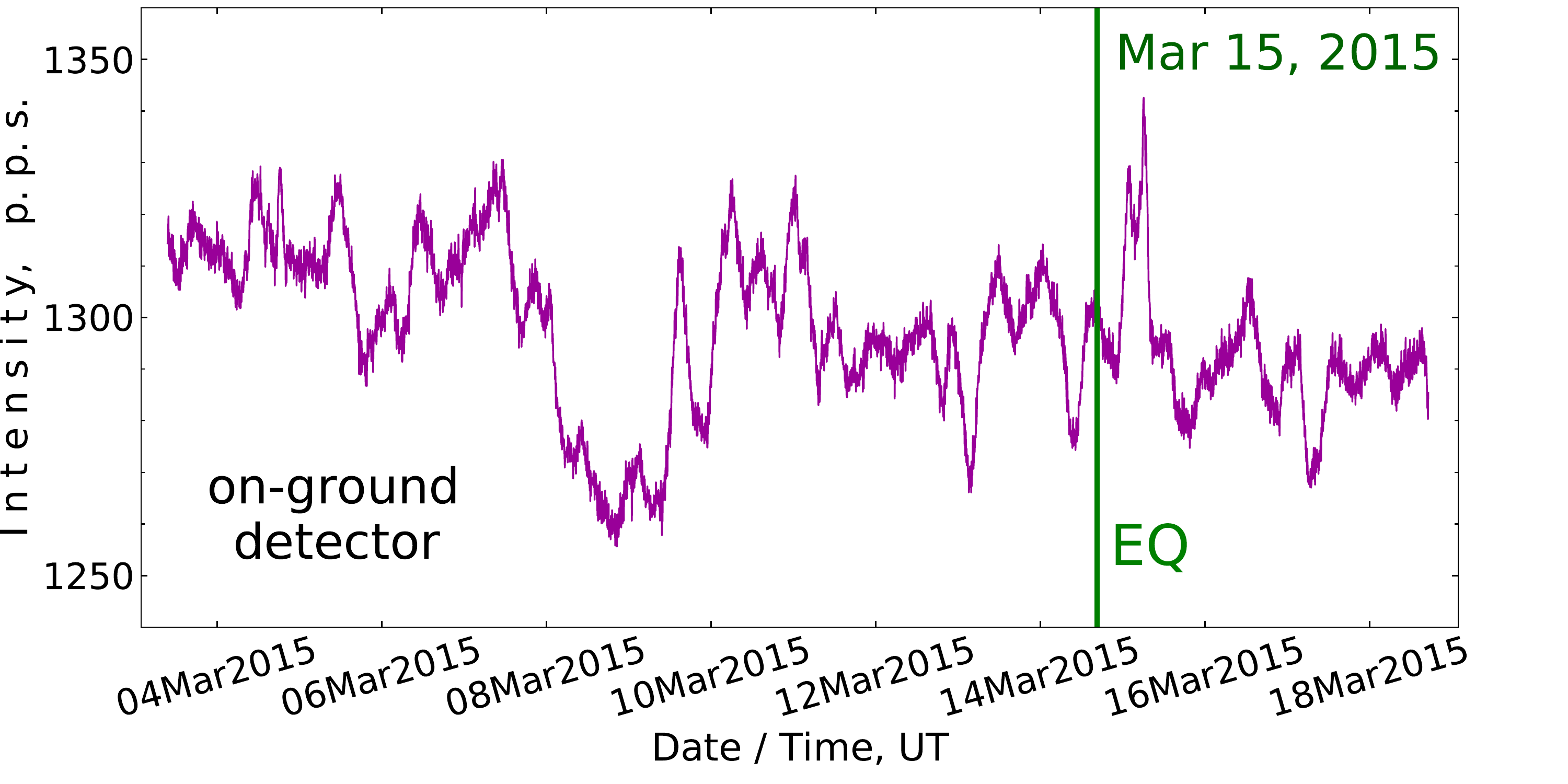}
\includegraphics[width=0.5\textwidth, trim=0mm 0mm 0mm 0mm]{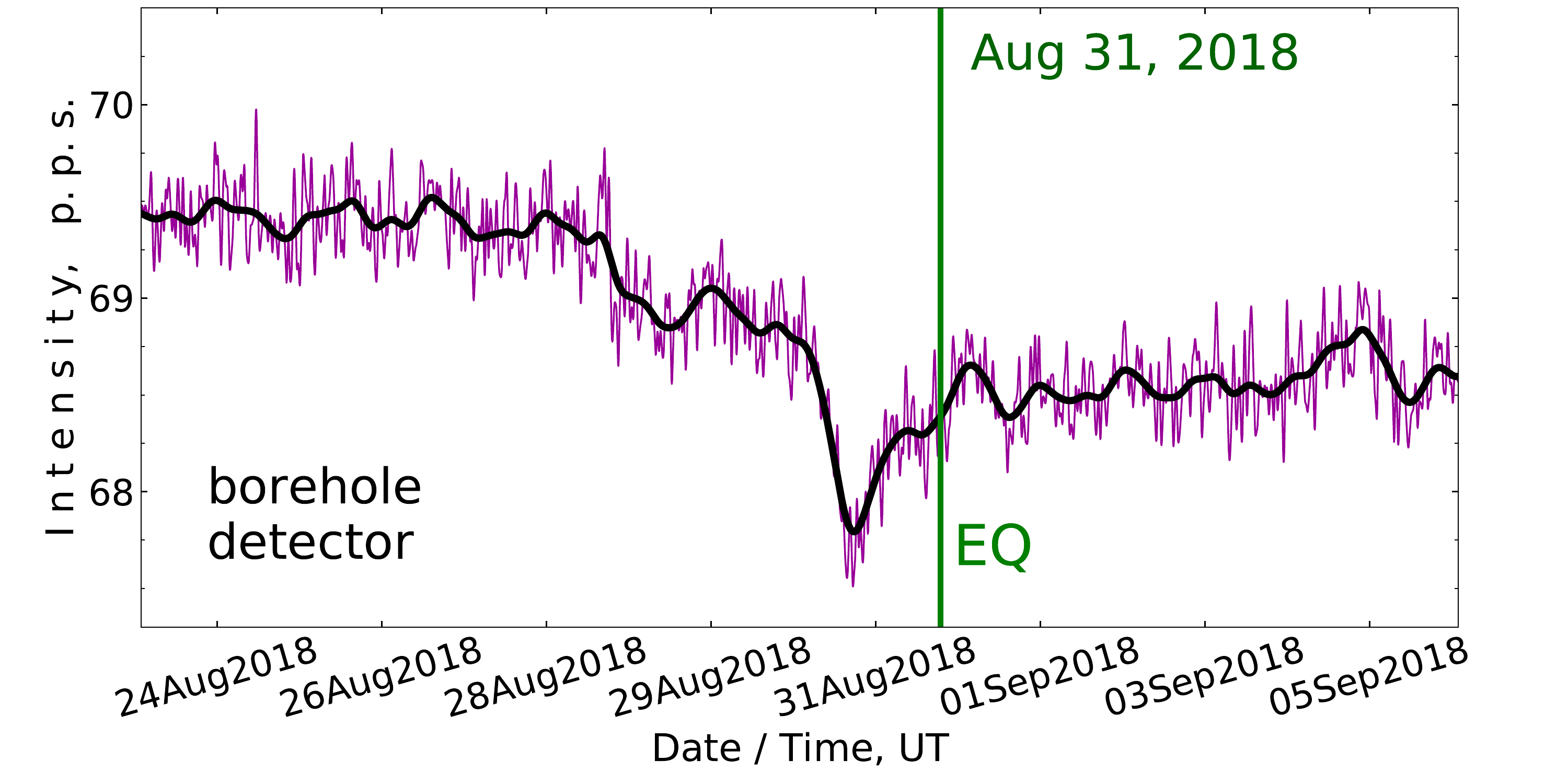}
\includegraphics[width=0.5\textwidth, trim=0mm 0mm 0mm 0mm]{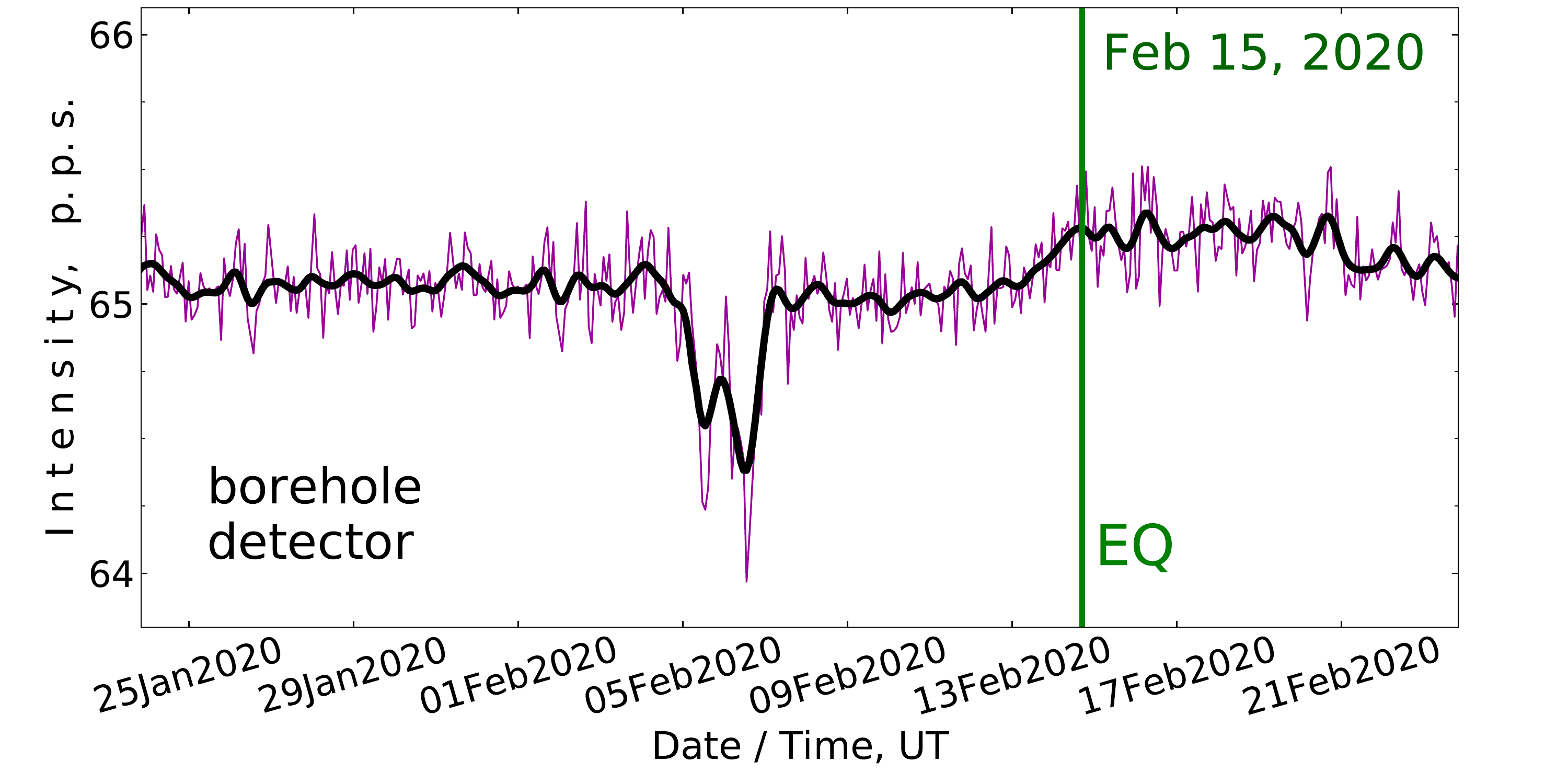}
\caption{
Examples of bay-like drops found in the smoothed records of gamma ray intensity before the earthquake events from Table\,\ref{tabitabi}. The precise earthquake moments are marked with vertical lines.
}
\label{figigammomoevents}}
\end{figure}

To enhance statistical reliability of the values in each distribution point and reduce the effect of random fluctuations it was made an averaging of the original time series: a mean of 60~neighbouring points in the distribution was calculated, so that the resulting series of the intensity measurements had the temporal resolution of $60\times10$\,s, and the time coordinates of its points were set in the middle of each of the subsequent 600\,s long time intervals. Afterwards, the averaged time series were additionally processed by the running average filter with the kernel length of 100~points
% (which is equivalent to the time duration of $6\cdot10^4$\,s, \ie $16.7$\,h)
%and with forced compensation of the --600\,s time shift  arising in the course of filtering.
.

%Данные у нас записываются один раз в десять секунд. Теперь делаем так, чтобы на графике данные отображались один раз в 600 с. Для этого берем интервал в 60 точек (600с), все значения по оси Y складываем и делим на 60 (определяем среднее). Точку на графике  ставим по середине этого интервала, т.е. сдвигаем на 30 точек влево по оси Х.

% Далее фильтруем  скользящим средним по 100 точкам (60000 с) со сдвигом на 1 точку (6000с) и чудесным образом получаем бухтообразное понижение вариаций гамма-квантов в скважине.  Этот эффект виден на 2 энергии (второй канал). На записях первой энергии никаких бухтообразных понижений не змечено. 

The smoothed distribution which appears after application of such averaging and filtering procedure is shown, piecewise, in the plots of Figure\,\ref{figigammomob}. As before, the values of the radiation intensity in this figure are expressed in the p.p.s. units, and only the data which correspond to the second amplitude channel with the energy threshold of detected gamma ray quanta of 150\,keV are considered here.
The time distribution which results from averaging of the original data is shown by the thin line in all three graphs, while the result of additionally applied filtering---by the thick black curve in the middle panel only. 
Among the three plot frames in Figure\,\ref{figigammomob} the upper one corresponds to the time period significantly preceding the moment of the January~19, 2020 earthquake, the middle one overlaps this moment, and the data acquired much later the earthquake are reproduced in the bottom plot. Each plot in the figure covers a prolonged time lapse of a few weeks order.

As it is seen in Figure\,\ref{figigammomob}, the smoothed intensity distribution remains rather uniform over the whole time span presented, with the only exception in the dates immediately preceding the earthquake event, where a characteristic bay-like drop is seen in the two curves of the middle panel, both the averaged and filtered ones. The duration of the time delay between the minimum of the deepening in the smoothed curves and the moment of the earthquake equals to about (7--8)\,days. It was found that most prominently the depression reveals itself in the distribution of the intensity of gamma ray quanta detected by the borehole detector with the energy threshold of 150\,keV; in the other energy ranges of the same detector the smoothing procedure fails to find any perceptible irregularities in that time.

In the course of long-term monitoring measurements at the Tien Shan station it was revealed that such bay-like drop with typical relative depth of (0.3--0.5)\% in the smoothed intensity records of (100--200) keV gamma rays is a rather characteristic phe\-no\-me\-non which was regularly reproduced in the time periods preceding the relatively close and powerful earth\-qua\-kes from Table\,\ref{tabitabi} (\ie those in which the station occurred within the Dobrovolsky distance). Sometimes, besides the intensity record of the borehole detector, it succeeds to find similar drop also in the data of the gamma detector which was operating above the surface of the ground  (see Section\,\ref{sectitechni}). A few other illustrative examples of the same kind are presented in Figure\,\ref{figigammomoevents}. In Table\,\ref{tabitabi} the presence or absence of the bay-like drop effect in the intensity records of the underground and surface detectors on the eve of an earthquake is marked in two special columns (the ``{\Large +}'' label means presence; since the detector in the borehole was installed only near the end of the year 2017, two leading cells in the corresponding column remain unfilled). It is seen that in all these cases anomalies in the smoothed intensity curves were found in the pre-earthquake time.

The most right column of Table\,\ref{tabitabi} contains  estimations of the time delay between the earthquake and the minimum point of the preceding bay-like drop. As it seen there, these delays vary between the limits of (2--8)\,days, with the mean of $\sim$5\,days, which makes the newly found effect of bay-like drops to be interesting from the viewpoint of timely earthquake forecast.
 
% mean days delay: (2+6+7+2+6+7+7.5+10)/8=5.9375
%>>> (2+6+7+2+6+8+7)/7 =  5.42

A bay-like drop in the time distribution of gamma radiation similar to those described here, together with a decrease in the flux of thermal neutrons in the near-surface atmosphere, was also observed at the Tien Shan station before the May\,1, 2011 M5.8 earthquake event; the results then acquired were reported in \cite{quakesour2012}.

\subsection{The close earthquake event of December~30, 2017}
On the date of December~30, 2017 a M4.2 earthquake has happened which had a focus location in nearest vicinity of the borehole at the Tien Shan mountain station, only $\sim$5.3\,km from the station, and at a $\sim$10\,km depth.
%below the surface of the earth's crust.
Such uniquely close position of the epicenter permitted to register in detail the anomalous effects in behaviour of various geophysical parameters which were detected in the time period preceding this event. The results of this study are presented in the following plots.

\begin{figure}
{\centering
\includegraphics[width=0.5\textwidth, trim=5mm 33mm 0mm 0mm]{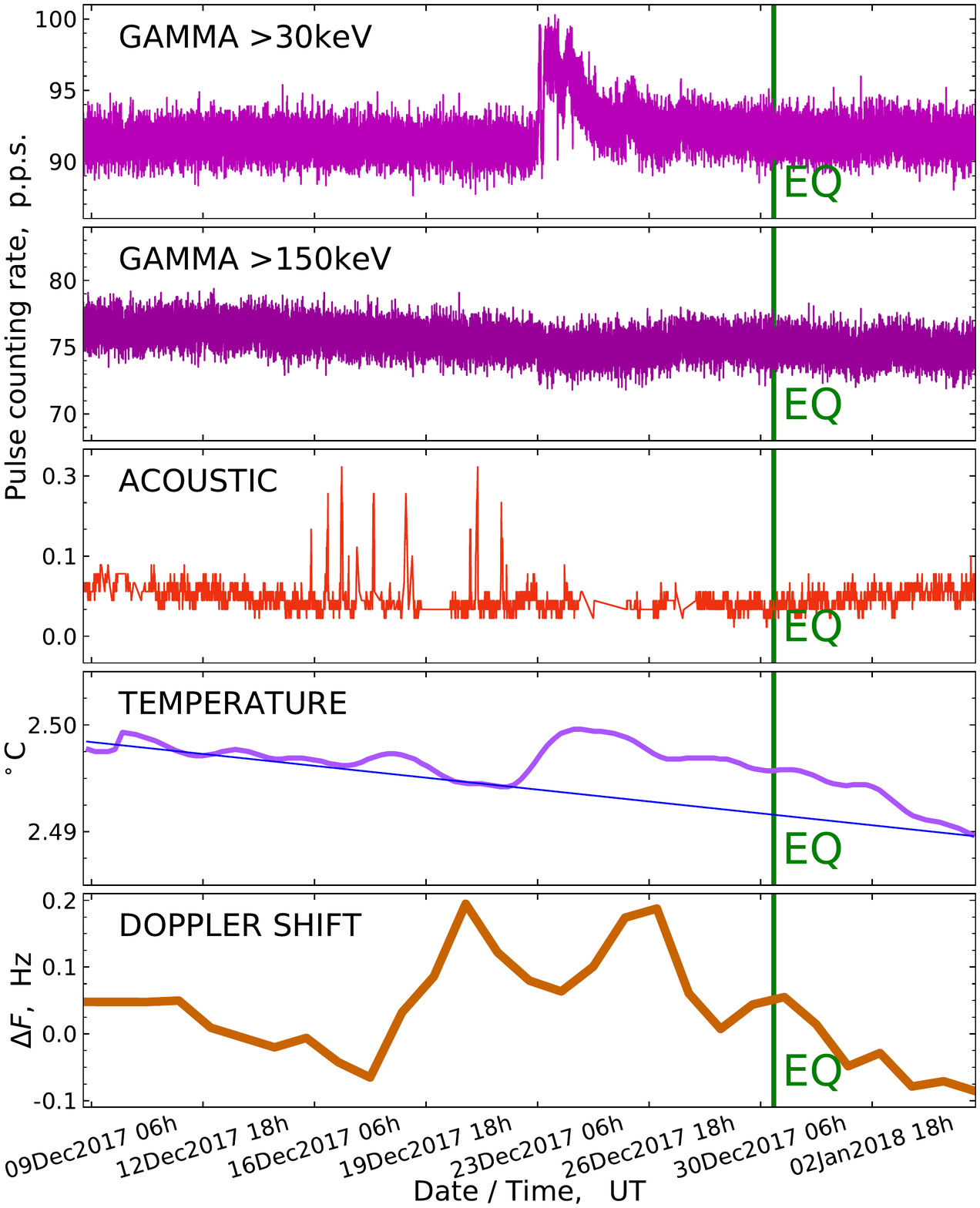}
\caption{The records of signal intensity at the detectors of the Tien Shan station which were written in the period preceding the December~30, 2017 earthquake. From top to bottom: the counting rate, in pulses per second, of the gamma ray signals detected in the $\geqslant$30\,keV channel of the underground detector, and of the high amplitude acoustic pulses; the temperature, in centigrade, measured at the depth of 40\,m, and the Doppler shift values. The moment of the earthquake (EQ) in each plot is marked by the thick vertical line.}
\label{figigammoskva}}
\end{figure}

% gamma in the hole
Figure\,\ref{figigammoskva} presents the data acquired at the earthquake time from the detectors operating in the borehole of the Tien Shan station. As it follows from the two upper frames of this figure, $\sim$7\,days before the earthquake an anomalous outburst of soft gamma rays was detected in the borehole, with its energy spectrum limited below 150\,keV, and with total duration achieving $\sim$2\,days. During the next 3\,days after the outburst the counting rate of low-energy gamma quanta remained noticeably disturbed. An enhanced level of the soft radiation intensity in the borehole was kept until a few days (2--3\,January) after the earthquake. The maximum excess of the gamma rays intensity at the peak of the detected outburst was $\sim$10\% above the preceding background level. With account to the area of the scintillator crystal, $\sim$75\,cm$^2$, and a $\sim$(30--50)\,\% registration probability of gamma ray quanta in the tens of keV energy range \cite{thunderour2016,thunderour2020touritge}, an absolute additional flux of the soft gamma radiation at the peak of the burst can be estimated as $\sim$(3--5)\,cm$^{-2}$s$^{-1}$.

% estimation of the relative excess:
%($\diameter$40$\times$40)\,mm$^2$.
% mean-before 913
% mean-after 923 (and 913 again at the very end)
% peak 1000
% an increase: (1000-913)/913 ~ 10%.

% the crystal area: 3.14*4*4 + 2*(3.14*2**2) ~ 75cm^2
% 100/75/0.3~1.3 

% gamma in the air
\begin{figure}
{\centering
\includegraphics[width=0.5\textwidth, trim=40mm 15mm 30mm 0mm]{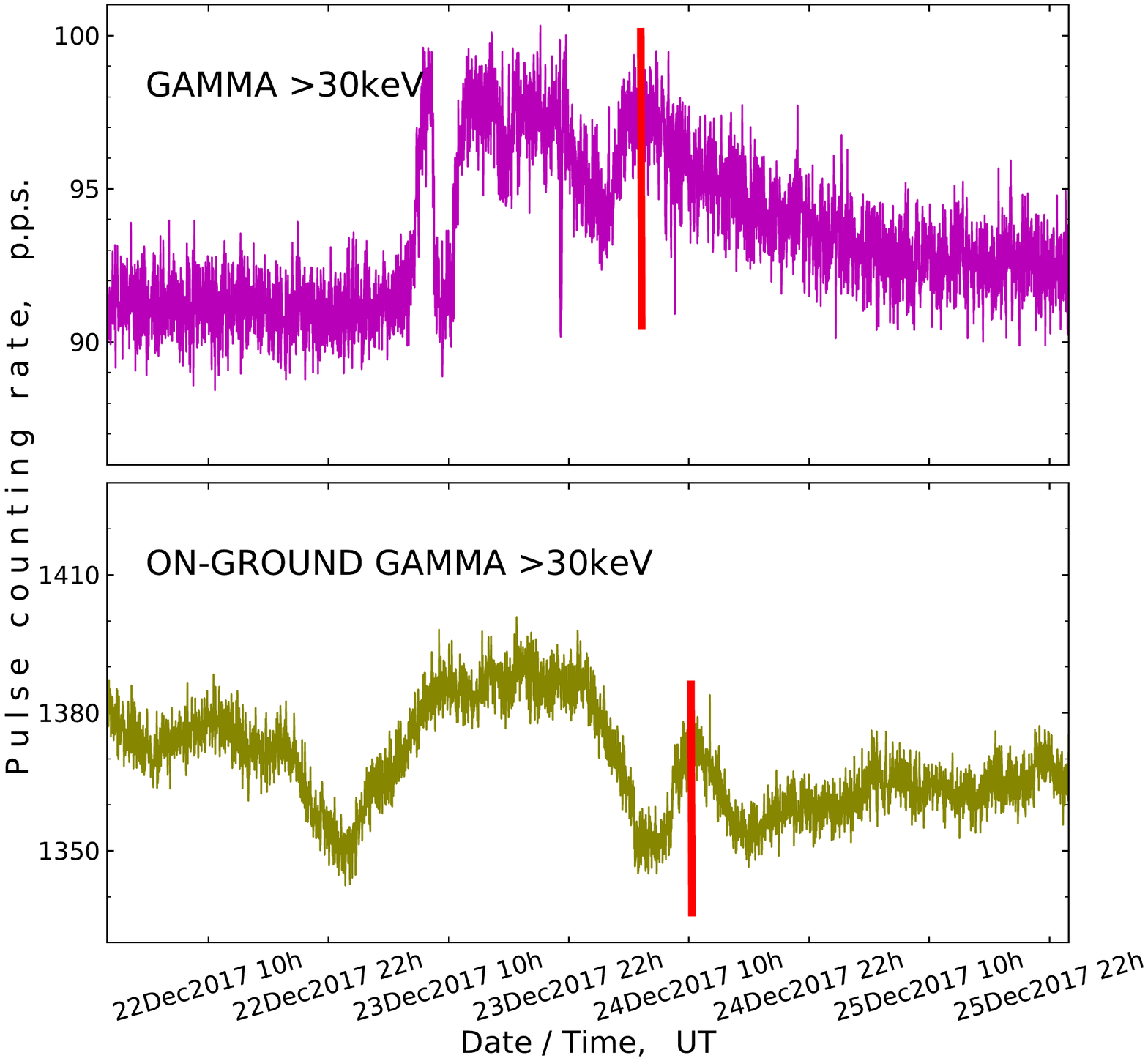}
\caption{Comparison of the gamma radiation intensity records made before the December~30, 2017 earthquake by the detectors placed both deep in the borehole (upper panel) and at the surface of the ground. Two mutually corresponding peaks of the counting rate of gamma rays are marked with vertical lines.
}
\label{figinazscinti}}
\end{figure}

Comparison of radiative response to the approaching earthquake between the data of the underground and surface gamma detectors is made in Figure\,\ref{figinazscinti}.
As it was mentioned above, in the near-surface atmosphere the radiation fluxes are largely influenced by precipitations and variation of the atmospheric temperature and pressure, but in the time of the gamma ray outburst on 23--24\,December any factors of that sort were absent.
% which prevents any unambiguous identification of the effects of seismic nature.
As it follows from Figure\,\ref{figinazscinti}, the time distribution of gamma ray intensity measured above the surface of the ground has much more smoothed shape than the time series of the borehole scintillator which permits to trace even the thin structure of the detected radiation disturbance. Besides, an increase of the radiation background in the near-surface atmosphere appears $\sim$5\,h later than the outburst of gamma rays in the borehole. Seemingly, this delay corresponds to the migration time of seismogenic radioactive elements from subsoil layers onto the surface.

% sanity check
\begin{figure}
{\centering
\includegraphics[width=0.5\textwidth, trim=0mm 0mm 0mm 0mm]{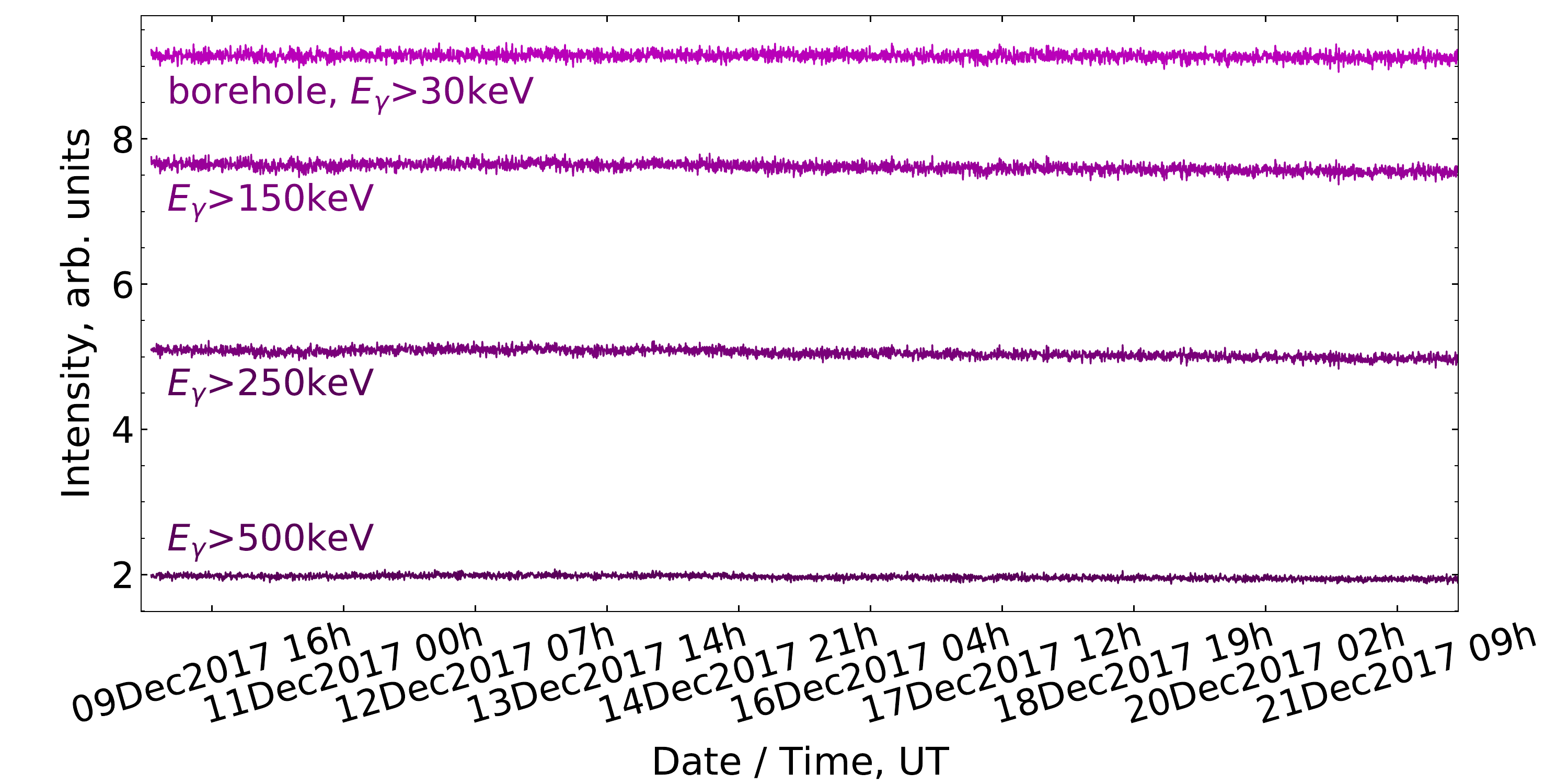}
\includegraphics[width=0.5\textwidth, trim=0mm 0mm 0mm 0mm]{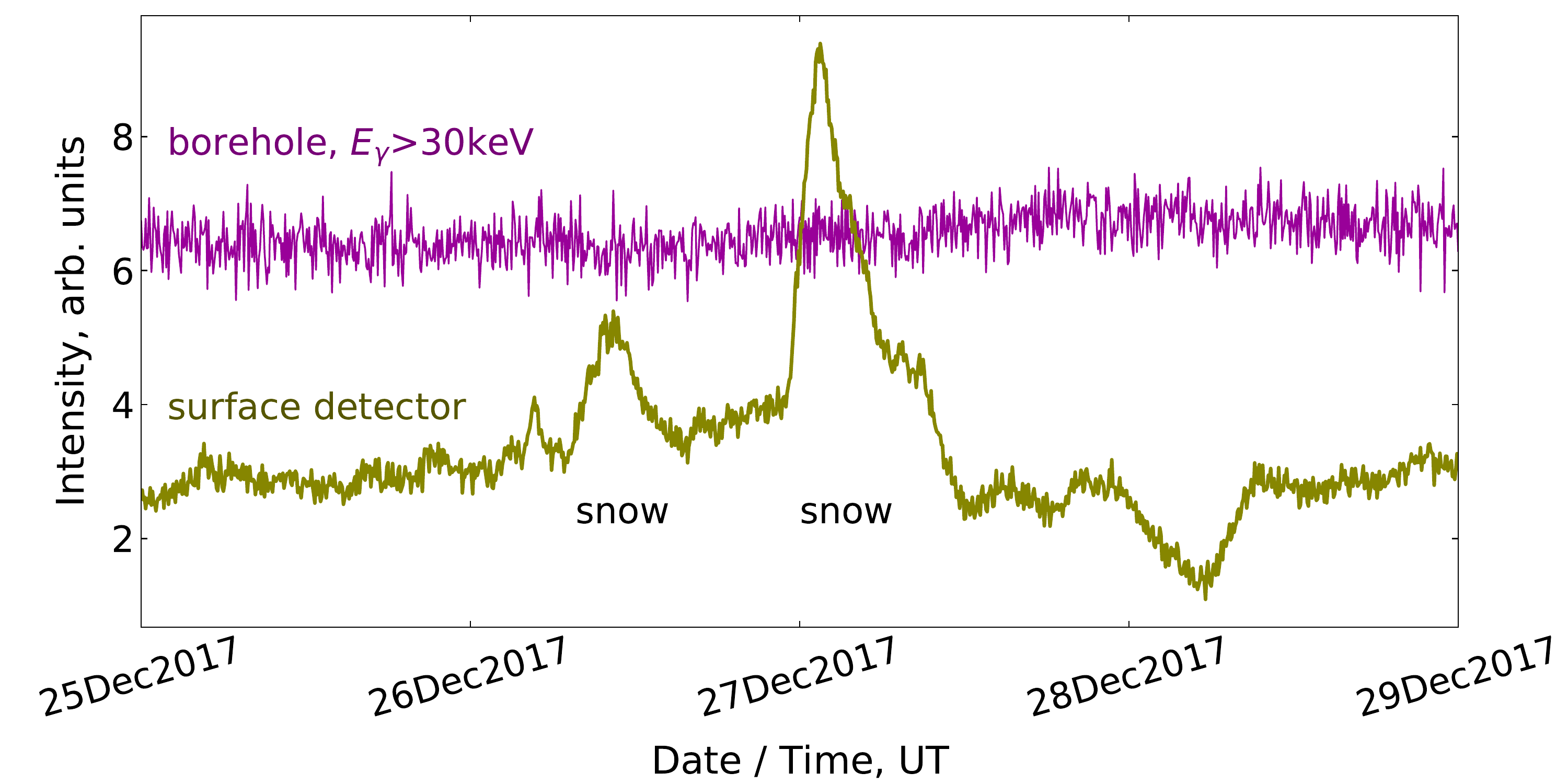}
\caption{Smoothed records of signal intensity in the gamma detectors written before (upper panel) and after (lower panel) the rise of soft radiation on 23--24 December 2017.}
\label{figigammobackgr}}
\end{figure}

In Figure\,\ref{figigammobackgr} the records of signal intensity in gamma detectors are shown which were written in the two time periods, immediately before and after the burst of soft radiation at 23--24 December 2017. The time distributions presented here were previously smoothed in accordance with the procedure described in Section\,\ref{sectisinu}. In the upper panel of Figure\,\ref{figigammobackgr} it is seen that before the outburst the radiation background in the borehole remained undisturbed. As it follows from the plot in the bottom panel, it was not influenced even in the periods of two intensive snowfalls which occurred on 26 and 27 December and leaved their imprints in the data of the surface detector. This data confirm reliability of the soft radiation increase in subsoil layers which was detected in the low-threshold channel of the borehole detector around the earthquake time.

% gamma power spectrum
\begin{figure}
{\centering
\includegraphics[width=0.5\textwidth, trim=0mm 0mm 0mm 0mm]{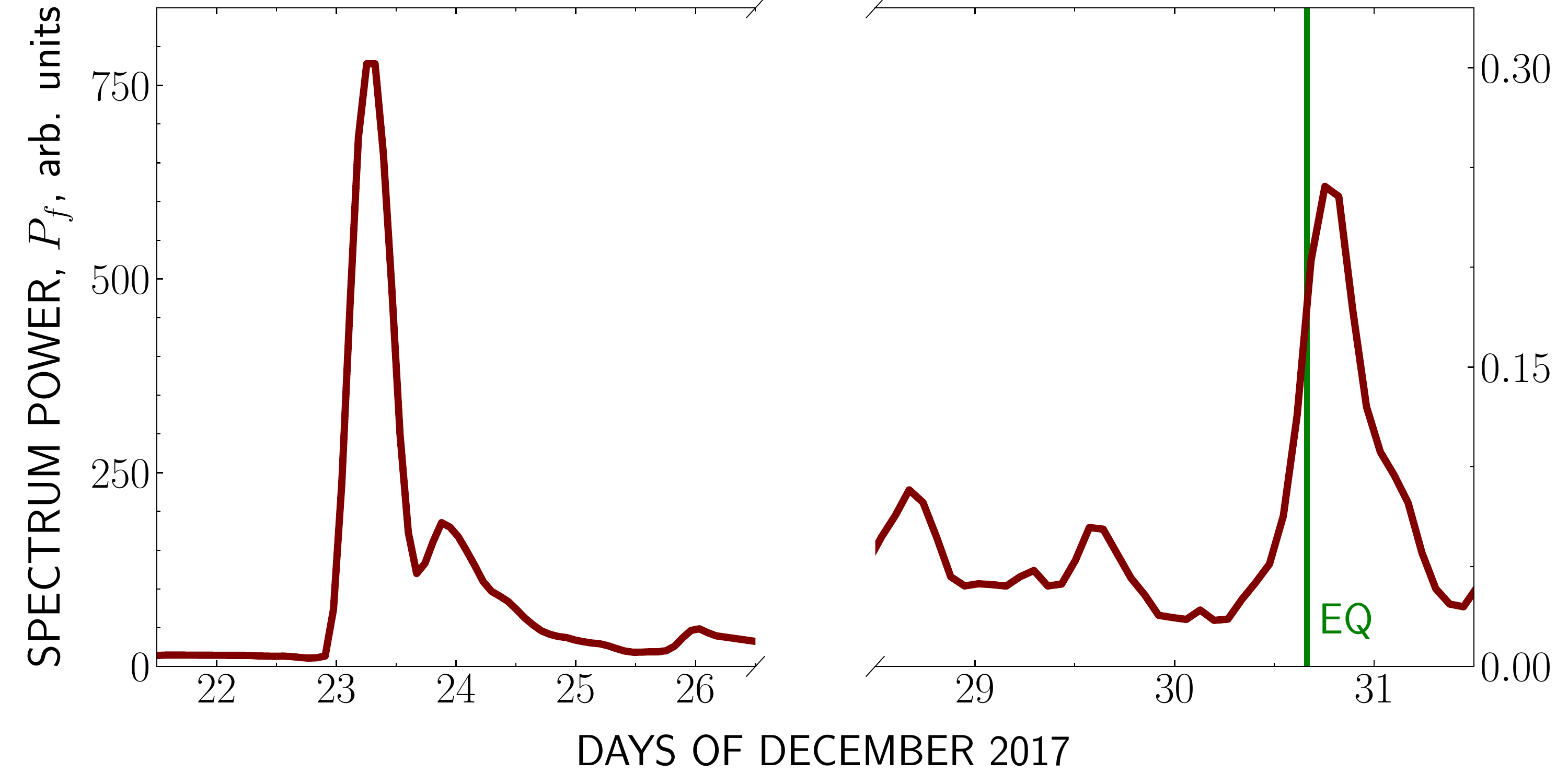}
\caption{The dynamic power spectrum of the gamma ray flux measured a few days before and after the December~30, 2017 earthquake. The vertical line marks the moment of the earthquake (EQ).}
\label{figispcpow}}
\end{figure}

Another approach to analyzing the behaviour of the gamma ray flux is based on calculation of the dynamic power spectrum of its variation. In contrast to the exploration of the time series applied above this method is capable to reveal the weak disturbances in a particular frequency range.

The calculation of the power spectrum of gamma radiation consists of the following. Among the time series of the counting rate in gamma detector it was selected a time window with a fixed duration, $T_w$, the frequency spectrum of the counting rate variation was defined within this window, and an integral average power $P_f$ of this spectrum was calculated between the frequency limits $[f_{low},f_{high}]$. Then the time window was displaced to the step $0.5\cdot T_w$, and the whole procedure was repeated. The parameters accepted for this algorithm in the present study were $T_w=6171$\,s ($1.7$\,h), $f_{low}=10^{-4}$\,s$^{-1}$ (or $1/2.8$\,h$^{-1}$), and $f_{high}=2.5\cdot 10^{-4}$\,s$^{-1}$ (or $1/1.1$\,h$^{-1}$).

In Figure\,\ref{figispcpow} it is presented the resulting distribution of the integral power values $P_f$ of the dynamic spectra calculated in the way described for the $E_\gamma\geqslant 30$\,keV channel of the underground gamma detector over a few days around the December~30, 2017 earthquake. As it follows from the left plot of this figure, at the peak of the gamma ray outburst on 23--24~December the integral power of the dynamic variation spectrum was $\sim$8000~times above the background. A much smaller, but noticeable increase of the spectrum power, 2.7~times above the mean square level of its background fluctuation, is seen also in the right plot of the same figure both before and after the earthquake moment. According to these high sensitivity data, the gamma radiation flux still remained disturbed during the whole next day after the seismic event.

% acoustic
The {\it ACOUSTIC} panel of Figure\,\ref{figigammoskva} demonstrates the time history of the counting rate of short time pulses with a peak amplitude above the threshold ADC code of 300 in the smoothed envelope channel of the underground microphone (see the description of the acoustic detector in Section\,\ref{sectitechni}). The one second normalized counting rate values presented here were obtained by averaging of the original pulse number measurements over successive 10\,min long time intervals.

It is seen in the {\it ACOUSTIC} panel that $\sim$7\,days before the growth of gamma radiation (and $\sim$10\,days before the earthquake) a quasi-regular sequence of the peaks of acoustic emission was detected in the borehole. The duration of each of the peaks was about (1--2)\,h, and the separating time gaps between them varied in the limits of (5--12)\,h. The sequence of acoustic pulses terminates shortly before the beginning of the radiation increase. %Outside the 10\,days long earthquake preceding period there was not observed any anomalies of similar quasi-regular kind.

% temperature
The data of precision temperature monitoring with the thermometer placed 40\,m below the surface of the ground are shown in the {\it TEMPERATURE} panel of Figure\,\ref{figigammoskva}. As it is seen there, just at the beginning of the seismic activity period there was registered in the borehole a noticeable temperature deviation up to $(5-7)\cdot10^{-3}$\,centigrade above the general cooling trend of the continuing winter season which is indicated in the plot by the thin straight line. Starting together with the rise of gamma ray intensity, the anomalous temperature enhancement terminates a few days after the earthquake. %Possibly, a smaller but perceptible temperature increase was also present in the time of the acoustic pulses sequence mentioned above.

% doppler
The Doppler shift measurements of the reflected radio wave at ionosphere sounding with the frequency 5121\,kH are presented in the bottom panel of Figure\,\ref{figigammoskva}. In this plot it is seen a significant drop of Doppler frequency which coincides with the beginning of the gamma ray outburst on 23--24\,December 2017. It should be noted that during the whole period of 18--31\,December the geomagnetic environment remained quiet, and there were absent any magnetic storms which could influence the ionosphere \cite{texix}. The geomagnetic quietness at that time was an important condition for the revealing of seismogenic effect in the ionosphere.

\begin{figure}
{\centering
\includegraphics[width=0.5\textwidth, trim=5mm 0mm 0mm 0mm]{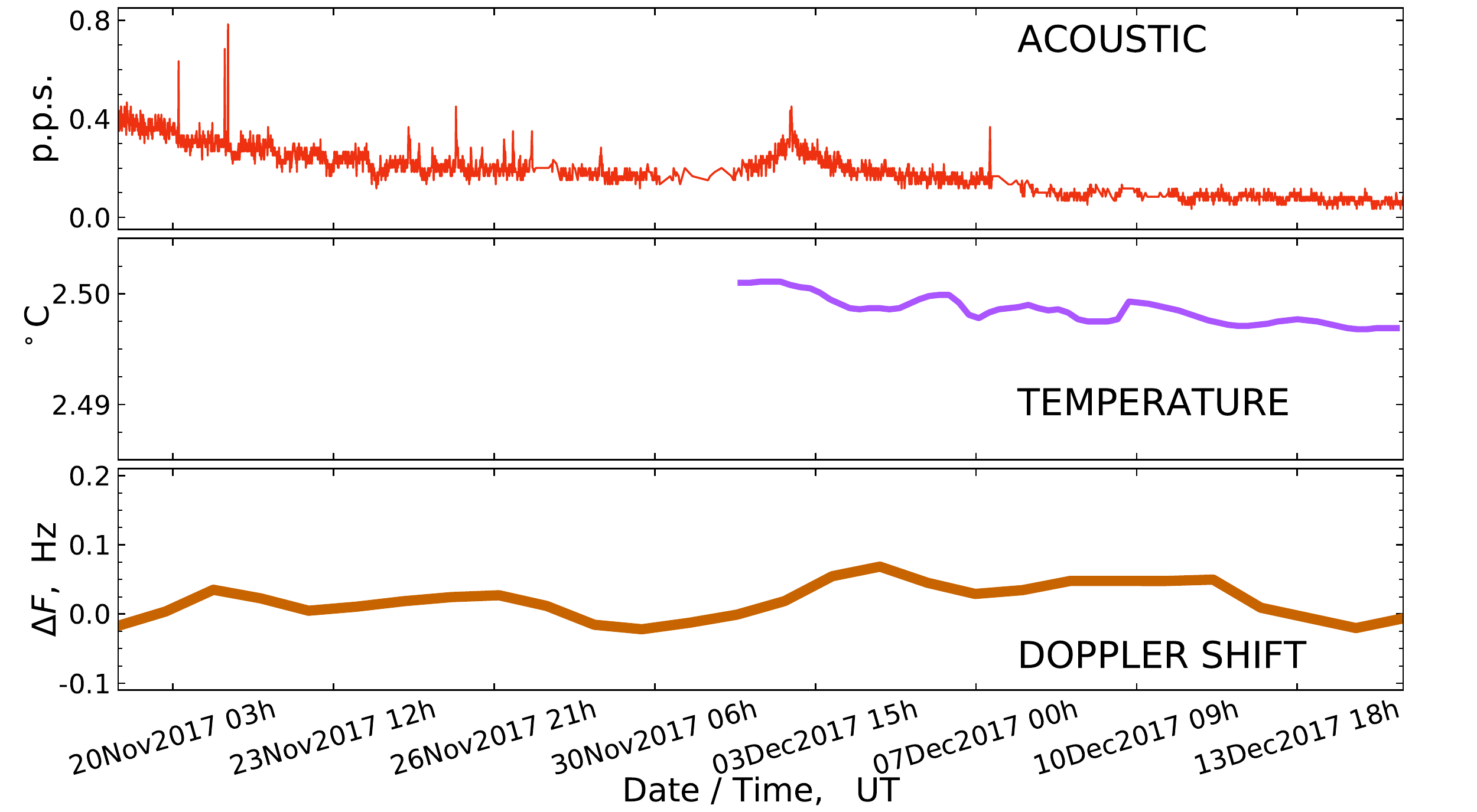}
\caption{Variation of the intensity of acoustic pulses and  temperature in the borehole, and of the Doppler shift of ionospheric signal during November---December 2017.}
\label{figibackgracoudoplitempli}}
\end{figure}

In Figure\,\ref{figibackgracoudoplitempli} the time history of the intensity of acoustic pulses and temperature in the borehole, as well as of the Doppler shift values is represented over a 3\,weeks long time lapse before the December~30, 2017 earthquake. (The temperature measurement in the borehole was started only on 1~December, so the plot in the corresponding panel of this figure is not complete). It is seen that outside the 10\,days period which has immediately preceded the earthquake there was not detected neither any anomaly of quasi-regular kind in the {\it ACOUSTIC} time series, nor significant deviation of the temperature and Doppler shift values from their usual uniform trends. This observation confirms the connection fidelity of the anomalies plotted in Figure\,\ref{figigammoskva} with the  seismogenic processes which were going on in the pre-earthquake time.

\section{Discussion}
Ionization of the near-surface atmosphere before earthquake plays the leading role in propagation of seismogenic disturbances from the lithosphere to ionosphere \cite{introref22,pulinets2015}. The present study is aimed to investigation of the intensity variations of ionizing gamma rays in the subsurface rock layers and in the near-surface atmosphere. One of the most informative geophysical markers of earthquake preparation is emanation of radon at the change of stressed-deformed state of rocks in a seismically active region \cite{introref1,introref2,introref3,introref7, introref4,introref5,introref6}.  In the present study for searching for the anomalous effects connected with radon emanation it was used the detection method of the gamma rays emitted by the daughter products of radon decay.

The intensity of gamma radiation, as well as radon concentration in the near-surface atmosphere are influenced by a variety of meteorological factors \cite{rainsour2009,introref16,introref14} which circumstance hinders to selection of seismogenic effects. To avoid the influence of precipitations and temperature variation the gamma detector used in the present study was placed in a borehole, at the depth of 40\,m. The observations have shown that the background flux of gamma rays in the borehole varies negligibly between the days, and, in contrast to the gamma radiation in the atmosphere, it is not influenced by the precipitations like rain, snow, and hail. This determines the application possibility of the equipment installed in the borehole for revealing of small and short-time changes in the gamma ray flux.

% bay-like things.
In the monitoring data of gamma radiation acquired during the times shortly preceding earthquakes there were detected two types of anomalies. The first is appearance, (2--10)\,days before an earthquake, of the bay-like drops in  time series of the gamma radiation flux, both measured in the borehole and in the near-surface atmosphere. In these events the epicenter of a M5.0--M6.2 earthquake occurred at a (15--354)\,km distance from the borehole, within the limits of Dobrovolsky radius \cite{dobrovolsky}.

Time behaviour of the intensity of detected gamma rays depends on the content of radon in the sub-soil layer of rocks. According to \cite{introref6}, the variation of radon concentration on the surface reflects the changes of the compression or extension strain in the deep lithosphere layers. Penetrability of a rock massive for gas fluids is defined by the presence or absence of internal mutually connected cracks and pores. Within the compression zone of a seismically active region of the earth's crust it is typical the closure of such cracks and pores with drop of radon content on the surface, while in a zone of expansion both the fracturing and porosity grow which leads to increase of radon emanation. In \cite{introref6} it is emphasized essential non-linearity in dependence of the speed of radon exhalation on the tension state of the crust deformation.
At Northern Tien~Shan the earthquakes take place at the condition of prevailing compression, under the action of squeezing forces of the thrusting Arabian and Hindustan tectonic plates.
Seemingly, the bay-like drops of the intensity of soft gamma rays detected  within the Dobrovolsky distance before the earthquakes from Table\,\ref{tabitabi} were originated as a result of rocks compression in zone of the earthquake preparation.

% gamma burst
The second type of anomalous effects was observed on the eve of a M4.2 earthquake on 30~December 2017. This was a unique event in which the focus of earthquake was located practically beneath the observation point, at a distance of 5.3\,km from the borehole. It was detected an outburst of soft gamma rays with an amplitude multiply above the level of background fluctuations which occurred 7\,days before the main shock of the earthquake. The outburst of gamma rays was accompanied by simultaneous growth of geoacoustic emission and a rising trend of the temperature in the borehole which evidence in favour of seismogenic origin of the detected anomalies.

Significant increase of ionizing radiation both in the borehole and, after a $\sim$5\,h long delay, above the soil surface can seemingly result from intensified exhalation of radon from the deep lithosphere layers in the near-epicenter region. A rise of soil radon concentration followed by an increase of gamma ray intensity a few days before an earthquake in Eastern Taiwan was reported in \cite{introref10,introref13}.

% co- and post-
In publications \cite{ciotoli2007,ciotoli2014} it is noted that anomalous changes in the content of radioactive substances may exist relatively long time after earthquake which is a good marker of an active geological fault. Indeed, if to consider the behaviour of the gamma ray flux during some days after earthquake in Figures\,\ref{figigammomob}--\ref{figigammoskva}, it can be found that frequently there was absent any returning trend to the level of background intensity which has been existing before the period of seismic activity. As illustration, both at the moment of the December~30, 2017 earthquake and during some days after it   the gamma ray background occurred to be disturbed, which effect reveals best in the dynamic power spectrum of the gamma intensity variation in Figure\,\ref{figispcpow}. Similarly, a disturbance was observed at the Tien~Shan station in the flux of environmental gamma rays and neutrons after the May~1, 2011 earthquake \cite{quakesour2012}.
As for the more distant earthquakes listed in Table\,\ref{tabitabi}, in the cases of the August~31, 2018 M5.0, and, especially, of the January~19, 2020 M6.2 earthquakes the flux of gamma rays remained suppressed (5--10)\,days since the main shock, and during some days after the February~15, 2020 M5.6 earthquake there was observed a tendency to an enhanced intensity of the radiation flux (see Figures\,\ref{figigammomob} and~\,\ref{figigammomoevents}). Seemingly, delayed restoration of the radiative background after an earthquake and the direction of the radiation intensity change reflect regional peculiarities of the   exhalation process of radon in the co- and post-seismic time.

% acoustic & temperature
Anomalous peaks of the intensity of acoustic signals in the pre-earthquake time, as plotted in Figure\,\ref{figigammoskva}, can result from intensification of the processes of micro-cracks formation within the stressed rock medium at the depth of lithosphere. It is known that such processes are accompanied by generation of elastic vibrations which propagate through the earth’s crust in the form of sound wave, called geoacoustic emission \cite{acou_annalen_2002,acou_lei_2014}. A $\sim$7\,days long time delay between the beginning of the sequence of anomalous increases in the counting rate of acoustic pulses and the rise of gamma radiation in the borehole may reflect the time which was necessary for the radioactive substances to penetrate over the micro-cracks newly opened  from the depth of the earth's crust into the subsoil layers. The locally increasing trend of the temperature in the borehole which generally coincides with the rise of gamma ray background can be also connected with exhalation of the deep lithosphere gases.

% doppler again
Simultaneously with the complex of anomalous effects observed before the December 30, 2017 earthquake both in the borehole and above the surface of the ground, an ionosphere disturbance was registered as well, which revealed itself through lowering of the Doppler shift of ionospheric signal.
% of the frequency of  reflected signal at ionosphere sounding.
%
There are two basic models of disturbance propagation from the lithosphere up to the ionosphere height which are presently discussed in the scientific literature: the acoustic mechanism and propagation through the intermediary electric currents and fields.

The acoustic mechanism is studied rather well. According to this model, the vertical movement of the earth's surface at an earthquake time generates infrasonic waves which achieve the ionosphere and can be registered in the Doppler shift of ionospheric signal \cite{introref29,introref27}.

The mechanism of disturbance propagation through the electric currents and fields is based on the concept of Global Electric Circuit suggested in 1921 by C.\,Wilson. Primary consequence of the stressed deformation change in the rocks before an earthquake is emanation of radioactive radon and the daughter products of its decay which leads to ionization of the near-surface atmosphere layers. According to the concept of li\-tho\-sphe\-re-at\-mo\-sphe\-re-iono\-sphe\-re coupling \cite{introref22,pulinets2015}, under the influence of ionization the conductivity of the atmospheric boundary layer changes next, which stimulates generation of the electric fields and formation of inhomogeneities of electron concentration in the ionosphere. In the frames of this concept the disturbance of the Doppler frequency on the eve of the December~30, 2017 earthquake can be considered as an ionosphere response of seismogenic origin.
Simultaneous detection  before the earthquake of the anomalies in different geophysical fields increases  both reliability of every particular observation and fidelity of its connection with the process of the earthquake preparation. 

% nuclear explosion
\begin{figure}
{\centering
\includegraphics[width=0.5\textwidth, trim=0mm 0mm 0mm 0mm]{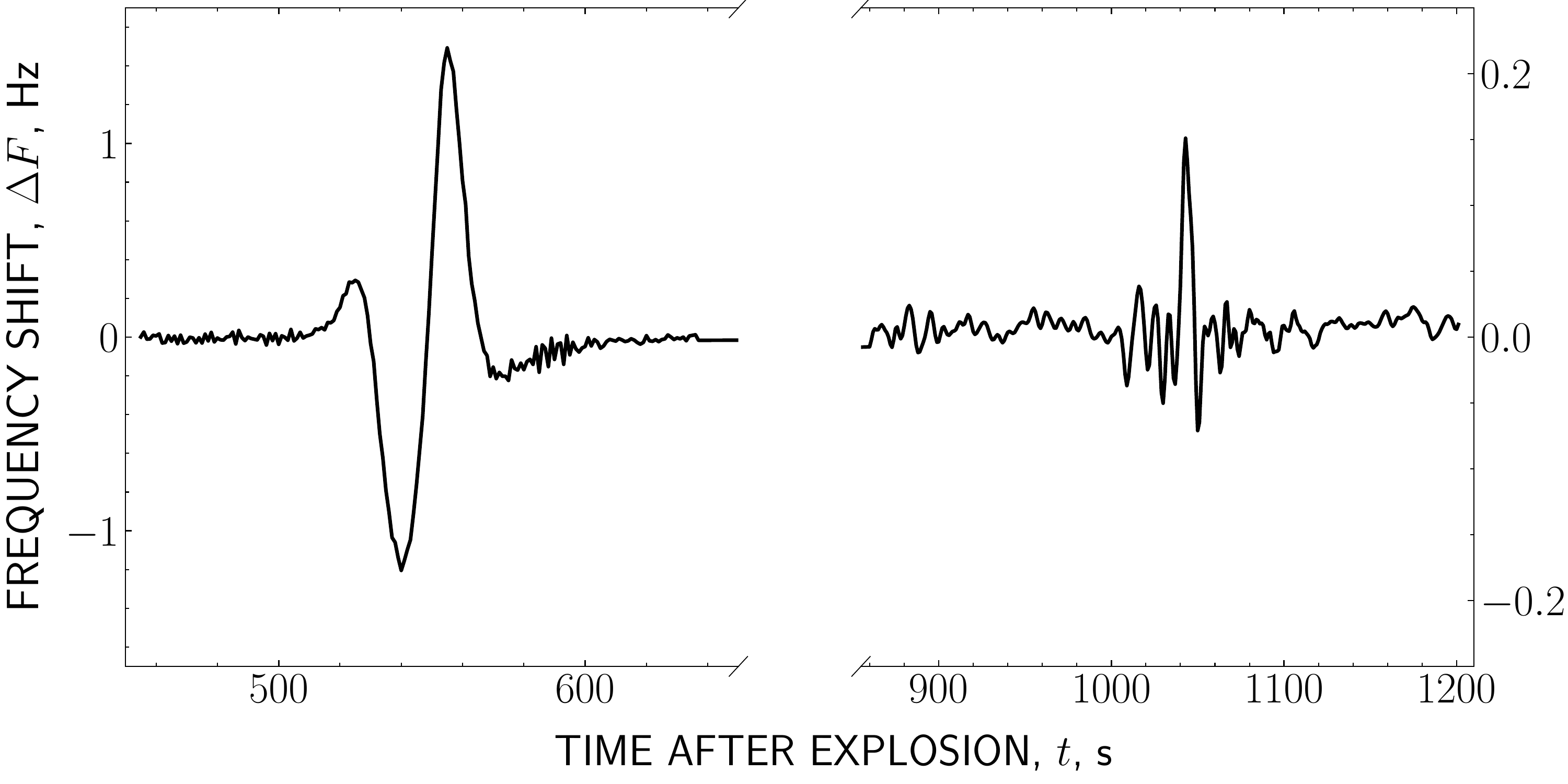}
\caption{The effect of Doppler frequency shift detected by ionosphere sounding after an underground nuclear explosion.}
\label{figiexplo}}
\end{figure}

A confirmation to the mechanism of disturbance propagation into the ionosphere stimulated by the rise of radioactivity in the near-surface atmosphere was obtained at a retrospective analysis of the Doppler shift records which were done in the times of underground nuclear explosions in 1980s at the Semipalatinsk testing site in Kazakhstan.
In Figure\,\ref{figiexplo} an original time history of Doppler shift values is presented which was acquired on October~19, 1989 after a 75\,kt TNT equivalent explosion. The Doppler ionosonde was placed at the testing site, such that the radio wave reflection point in the ionosphere situated immediately above the epicenter.
It is seen in Figure\,\ref{figiexplo} that 510\,s (8.5\,min) since the explosion a large perturbation appears in the time series of Doppler frequency caused by penetration of the intensive acoustic wave into the ionosphere. The source of this wave is the movement of the earth's surface within the splitting zone of the explosion. Later on, at the time mark of 1005\,s (16.75\,min) another, essentially weaker, disturbance is present in the Doppler shift record.

The plot of Figure\,\ref{figiexplo} is one of the records made in the times of underground nuclear explosions at the Semipalatinsk testing site, when according to the data of dosimetric service the radioactivity level in the near-surface atmosphere above the epicenter was enhanced up to (15--40)\,R/h. 
It was found that secondary disturbances of smaller amplitude were frequently met in such records. Usually these disturbances appeared with a (15--18)\,min time delay, at an exit onto the earth's surface of the radioactive components of the explosion, and of the large masses of natural radon and thorium ejected by the seismic influence of the explosion on the rock.
It is interesting to note that at a 260\,t TNT ground surface fugas explosion in 1981 there was not detected any secondary effect in the Doppler shift record \cite{explo1987}. For comparison, it can be mentioned the results of a modern analysis of the consequences of an underground nuclear explosion on February~12, 2013 (Pungery, Northern Korea)  \cite{pulinets2015}. After the explosion there was registered a M4.9 seismic shock, and appearance of a negative anomaly of electron concentration in the ionosphere was detected as well according to the data of GPS satellites. This confirms the supposition of the authors of \cite{pulinets2015} that the burst ionization at the exit of radioactive components of the underground explosion onto the surface can be accompanied by local increase of the atmosphere conductivity and dropping of electron concentration in the ionosphere above the region of enhanced conductivity.

Thus, in the current study it succeeded to trace experimentally an appearance of disturbances in the ionosphere under triggering influence of ionization growths in the near-surface atmosphere, both connected with an underground nuclear explosion, and with a M4.2 earthquake. While in the case of explosion the release of ionizing radioactive substances took place {\it after} the explosion, in the period of earthquake preparation the exhalation of radioactive substances (radon and its decay products) was going on {\it before} the main seismic shock. 
%
%On the basis of presented  data it can be supposed that the ionization in the near-surface atmosphere is a universal basic factor of the process of disturbance propagation among the lithosphere-atmosphere-ionosphere system.
%
The data presented here can be considered as direct experimental confirmation of the propagation of seismogenic disturbances from the lithosphere up to the ionosphere in agreement with the concept of li\-tho\-sphe\-re-at\-mo\-sphe\-re-iono\-sphe\-re coupling, in which the role of initiating link plays the ionization of the near-surface atmosphere.

\section{Conclusion}
The main results of the present investigation can be summarized as the following.

1. Presently complex monitoring of the various geophysical fields is going on at the Tien Shan mountain scientific station and at the ``Orbita'' radio-polygon which are located in a seismically hazardous region of Northern Tien Shan, in vicinity of the sources of disastrous Vernen (M7.3), Chilick (M8.3), and Kemin (M8.2) earthquakes. Main purpose of this activity is searching for the seismogenic effects which precede earthquake.

2. Before a M4.2 earthquake with close epicenter distance of 5.3\,km it was found simultaneous appearance of anomalous effects in the variation of geoacoustic emission, temperature, and gamma radiation in subsoil rock layers (at a depth of 40\,m in a borehole), as well as of the gamma rays in the near-surface atmosphere. The time delay between the beginning of the detected anomalies and the earthquake was of about (7--10)\, days.

3. Simultaneously with the said seismogenic anomalies the disturbances were found also in the Doppler shift of ionospheric signal on the low-inclined radio-pass. This effect is considered here in the frames of the lithosphere-atmosphere-ionosphere coupling concept as an ionosphere response of seismogenic origin.

4. Also, the ionosphere disturbances were detected by the Doppler shift of ionospheric signal at the rise of radioactivity in the near-surface atmosphere after  underground nuclear explosions.

5. Characteristic bay-like drops in the time series of the intensity of gamma radiation, both in the subsoil layers and in the atmosphere, were found (2--8)\,days before the M5.0--M6.2 earthquakes which had the epicenter location at the distances of (15--354)\,km from the gamma detector, within the limits of Dobrovolsky radius. 

\section*{Acknowledgments}
This research was funded by the Ministry of Education and Science of the Republic of Kazakhstan, grant number AP09260262 ''Monitoring and research of geosphere interactions in the lithosphere-atmosphere-ionosphere system in geodynamically active regions''.

%\newcommand{\url}[1]{{~#1}}
%\newcommand{\href}[1]{{~#1}}

%\section*{References}

\end{document}